\pdfoutput=1
\documentclass[12pt]{article}
\usepackage{graphicx}
\usepackage{subcaption}
\usepackage{amssymb}
\usepackage{amsmath}
\usepackage{amsfonts}
\usepackage{multirow}

\usepackage{xcolor}
\usepackage{url}
\usepackage{cancel}
\usepackage{ulem}
\usepackage{float}
\usepackage{float}

\usepackage{pifont}

\usepackage{soul}
\usepackage[english]{babel}
\usepackage[T1]{fontenc}
\usepackage{lmodern}
\usepackage[utf8]{inputenc}  
\usepackage{bbm}

\usepackage{hyperref}
\hypersetup{
  colorlinks   = true, 
  urlcolor     = blue, 
  linkcolor    = black, 
  citecolor   = black 
}

\usepackage[numbers,sort&compress]{natbib}
\graphicspath{{plots/}}

\newcommand{\lsim}
{\;\raisebox{-.3em}{$\stackrel{\displaystyle <}{\sim}$}\;}
\newcommand{\gsim}
{\;\raisebox{-.3em}{$\stackrel{\displaystyle >}{\sim}$}\;}

\newcommand\tb{\tan\beta}

\newcommand\ReDiag{\mathop{%
  \raise .5pt\hbox{[}%
  \widetilde{\mathrm{Re}}%
  \raise .5pt\hbox{]}}}
\newcommand\ReOffDiag{\mathop{%
  \raise .5pt\hbox{$\llbracket$}%
  \widetilde{\mathrm{Re}}%
  \raise .5pt\hbox{$\rrbracket$}}}

\newcommand\MZ{M_Z}
\newcommand\Mh{M_h}

\newcommand\MA{M_A}

\newcommand\Atau{A_\tau}

\newcommand\Sn{\tilde\nu}
\newcommand\Sl{\tilde l}

\newcommand\Slpm{\tilde l^\pm}

\newcommand\Sel[1]{\tilde e_{#1}}
\newcommand\Smu[1]{\tilde \mu_{#1}}

\newcommand\mse[1]{m_{\Sel{#1}}}
\newcommand\msl[1]{m_{\Sl_{#1}}}

\newcommand\Stau[1]{{\tilde\tau_{#1}}}
\newcommand\stau{\tilde \tau}
\newcommand\sntau{\tilde \nu_\tau}

\newcommand\mL{m_{\tilde l_L}}
\newcommand\mR{m_{\tilde l_R}}

\newcommand\msnutau{m_{\tilde \nu_\tau}}

\newcommand\ino[1]{\tilde\chi_{#1}}

\newcommand\chapm[1]{\ino{#1}^\pm}
\newcommand\champ[1]{\ino{#1}^\mp}
\newcommand\chap[1]{\ino{#1}^+}
\newcommand\cham[1]{\ino{#1}^-}
\newcommand\cha{\chapm}
\newcommand\mcha[1]{m_{\chapm{#1}}}

\newcommand\neu[1]{\ino{#1}^0}
\newcommand\mneu[1]{m_{\neu{#1}}}

\newcommand\refeq[1]{Eq.~(\ref{#1})}
\newcommand\refeqs[1]{Eqs.~(\ref{#1})}

\newcommand\refse[1]{Sect.~\ref{#1}}

\newcommand\citere[1]{Ref.~\cite{#1}}
\newcommand\citeres[1]{Refs.~\cite{#1}}

\newcommand{\CP}{{\cal CP}}
\newcommand{\cp}{{\CP}}

\newcommand{\tev}{\,\, \mathrm{TeV}}
\newcommand{\gev}{\,\, \mathrm{GeV}}
\newcommand{\mev}{\,\, \mathrm{MeV}}

\newcommand\MO{\texttt{MicrOMEGAs}}
\newcommand\CM{\texttt{CheckMATE}}

\newcommand\pb{\ensuremath{\,\mbox{pb}}}
\newcommand\fb{\ensuremath{\,\mbox{fb}}}
\newcommand\ab{\ensuremath{\,\mbox{ab}}}

\newcommand\ifb{\ensuremath{\,\fb^{-1}}}
\newcommand\iab{\ensuremath{\,\ab^{-1}}}

\newcommand\msmu[1]{m_{\tilde{\mu}_{#1}}}
\newcommand\mstau[1]{m_{\tilde{\tau}_{#1}}}

\newcommand{\sig}{\sigma}

\def\order#1{\ensuremath{{\cal O}(#1)}}
\def\reffi#1{\mbox{Fig.~\ref{#1}}}

\def\ga{\gamma}
\def\De{\Delta}

\def\gmin2{\ensuremath{(g-2)_\mu}}
\def\amu{\ensuremath{a_\mu}}
\def \met  {\mbox{${E\!\!\!\!/_T}$}}
\newcommand{\ssi}{\ensuremath{\sig_p^{\rm SI}}}

\definecolor{Orange}{named}{orange}
\definecolor{Purple}{named}{purple}
\definecolor{Lightblue}{cmyk}{0.9,0.1,0.1,0.3}
\definecolor{dgelborange}{cmyk}{0.,0.3,0.5, 0.}
\definecolor{Lila}{rgb}{0.5,0.,1}
\definecolor{Darkgreen}{rgb}{0.,.7,0.2}

\evensidemargin \oddsidemargin
\allowdisplaybreaks
\begin{document}
\thispagestyle{empty}

\def\thefootnote{\fnsymbol{footnote}}

\begin{flushright}
\mbox{}
IFT--UAM/CSIC--23-080 
\end{flushright}


\begin{center}

{\large\sc 
{\bf \boldmath{\gmin2} and Stau coannihilation:\\[.5em]
Dark Matter and Collider Analysis}}

\vspace{0.3cm}

{\sc
Manimala Chakraborti$^{1}$%
\footnote{email: M.Chakraborti@soton.ac.uk}%
, Sven Heinemeyer$^{2}$%
\footnote{email: Sven.Heinemeyer@cern.ch}%
and Ipsita Saha$^{3}$%
\footnote{email: ipsita@iitm.ac.in}
}

\vspace*{.5cm}

{\sl
$^1$School of Physics and Astronomy, University of Southampton,
\\ Southampton, SO17 1BJ,
United Kingdom.

\vspace*{0.1cm}

$^2$Instituto de F\'isica Te\'orica (UAM/CSIC), 
Universidad Aut\'onoma de Madrid, \\
Cantoblanco, 28049, Madrid, Spain

\vspace*{0.1cm}

$^3$Department of Physics, Indian Institute of Technology Madras,\\
Chennai 600036, India
}

\end{center}

\vspace*{0.1cm}

\begin{abstract}
\noindent
Slepton coannihilation is one of the most promising scenarios that
can bring the predicted Dark Matter (DM) abundance
in the the Minimal Supersymmetric Standard Model (MSSM)
into agreement with the experimental observation.
In this scenario, the lightest supersymmetric particle (LSP),
usually assumed to be the lightest neutralino, $\neu1$ can serve as a Dark Matter (DM)
candidate while the sleptons as the next-to-LSPs (NLSPs) lie close in mass.
In our previous studies analyzing the electroweak sector of MSSM,
a degeneracy between the three generations of sleptons was assumed for the sake of simplicity.
In case of slepton coannihilation this 
directly links the smuons involved in the explanation for \gmin2\ to the
coannihilating NLSPs required to explain the DM content of the universe.
On the other hand, in well-motivated top-down models such degeneracy do not hold,
and often the lighter stau turns out to be the NLSP at the electroweak (EW) scale,
with the smuons (and selectrons) somewhat heavier.
In this paper we analyze such a scenario at the EW scale
assuming non-universal slepton masses  where  the first two generations of sleptons
are taken to be mass-degenerate and heavier than the staus, enforcing
stau coannihilation. We analyze the parameter space of the electroweak MSSM in the light of
a variety of experimental data namely, the DM relic density and direct detection limits, LHC data and
especially, the discrepancy between the experimental result for the
anomalous magnetic moment of the muon, \gmin2, and its Standard Model
(SM) prediction.  We find an upper limit on the lightest neutralino
mass, the lighter stau mass and the mass of the tau sneutrino of
about $\sim 550 \gev$. In contrast to the scenario with full
degeneracy among the three families of sleptons, the upper limit on the
light smuon/selectron mass moves up by $\sim 200 \gev$.
We analyze the DD prospects as well as the physics
potential of the HL-LHC and a future high-energy $e^+e^-$ collider to
investigate this scenario further. We find that the combination DD experiments
and $e^+e^-$ collider searches with center of mass energies up to $\sqrt{s} \sim 1100 \gev$
can fully cover this scenario.
\end{abstract}


\def\thefootnote{\arabic{footnote}}
\setcounter{page}{0}
\setcounter{footnote}{0}

\newpage


\section{Introduction}
\label{sec:intro}
One of the main objectives in today's
collider physics as well as ``direct detection'' (DD) searches
is to understand the nature and origin of Dark Matter (DM).
A leading candidate among the plethora of 
Beyond the Standard Model (BSM) theories that predict a
viable DM particle is the Minimal Supersymmetric (SUSY) Standard Model  
(MSSM)~\cite{Ni1984,Ba1988,HaK85,GuH86} (see \citere{Heinemeyer:2022anz}
for a recent review).
MSSM extends the particle content of the Standard
Model (SM) by predicting
two scalar partners for all SM fermions as well as fermionic partners to all SM bosons. 
Furthermore, contrary to the SM case, the MSSM requires the presence of two Higgs
doublets, resulting in five physical Higgs bosons instead of the single Higgs
boson in the SM, namely the light and heavy $\cp$-even Higgs bosons, 
$h$ and $H$, the $\cp$-odd Higgs boson, $A$, and a pair of charged Higgs
bosons, $H^\pm$.
The SUSY partners of the SM leptons and quarks are known as the scalar leptons
and quarks (sleptons, squarks) respectively.
The neutral SUSY partners of the neutral Higgses and electroweak (EW) gauge
bosons give rise to the four neutralinos, $\neu{1,2,3,4}$. The corresponding
charged SUSY partners are the charginos, $\cha{1,2}$.
In an R-parity conserving scenario of MSSM the lightest neutralino
can be the lightest SUSY particle (LSP), resulting in a good DM candidate.
Depending on its nature, it can make up the full DM content of the
universe~\cite{Go1983,ElHaNaOlSr1984}, or only a fraction of it.
In the latter case, an additional DM component is required,
which could be, e.g., a SUSY axion~\cite{Bae:2013bva},
to saturate the experimentally measured relic density.

In \citeres{CHS1,CHS2,CHS3,CHS4,gmin2-mw} we performed a comprehensive analysis
of the EW sector of the MSSM, taking into account all
relevant theoretical and experimental constraints.
The experimental results comprised in
particular the  deviation of the anomalous magnetic moment
of the muon (either the previous result~\cite{CHS1,CHS2}, or the new,
stronger limits~\cite{CHS3,CHS4})%
\footnote{
Other evaluations of \gmin2\ within the framework of SUSY using the new combined
deviation $\De\amu$ (see \refse{sec:constraints}) can be found
in \citeres{Endo:2021zal,Iwamoto:2021aaf,Gu:2021mjd,VanBeekveld:2021tgn,Yin:2021mls,Wang:2021bcx,Abdughani:2021pdc,Cao:2021tuh,Ibe:2021cvf,Cox:2021gqq,Han:2021ify,Heinemeyer:2021zpc,Baum:2021qzx,Zhang:2021gun,Ahmed:2021htr,Athron:2021iuf,Aboubrahim:2021rwz,Chakraborti:2021bmv,Baer:2021aax,Altmannshofer:2021hfu,Chakraborti:2021squ,Zheng:2021wnu,Jeong:2021qey,Li:2021pnt,Dev:2021zty,Kim:2021suj,Ellis:2021zmg,Zhao:2021eaa,Frank:2021nkq,Shafi:2021jcg,Li:2021koa,Aranda:2021eyn,Aboubrahim:2021ily,Nakai:2021mha,Li:2021cte,Li:2021xmw,Lamborn:2021snt,Fischer:2021sqw,Forster:2021vyz,Ke:2021kgy,Ellis:2021vpp,Athron:2021dzk,Chakraborti:2021ynm,Chapman:2021gun,Aboubrahim:2021myl,Antoniadis:2021mqz,Acuna:2021rbg,Ali:2021kxa,Djouadi:2021wvb,Wang:2021lwi,Wang:2022rfd,Chakraborti:2022sbj,Boussejra:2022heb,Dermisek:2022hgh,Cao:2022chy,Gomez:2022qrb,Ahmed:2022ude,Chatterjee:2022pxf,Chakraborti:2022vds,Agashe:2022uih,Athron:2022uzz,Endo:2022qnm,Chigusa:2022xpq,Shang:2022hbv,Hamaguchi:2022wpz,Du:2022pbp,Tang:2022pxh,Yang:2022gvz,Cao:2022htd,Athron:2022isz,Maselek:2022cjb,Li:2022zap,Dickinson:2022hus,Zheng:2022ssr,Dao:2022rui,Heinemeyer:2022ith,Domingo:2022pde,Cao:2022ovk,Datta:2022bvg,Yang:2022qyz,Wang:2022wdy,Zhao:2022pnv,Chakraborti:2022wii,Borah:2023zsb,Ajaib:2023jhc,Baum:2023inl,SalihUn:2023unj,He:2023lgi,Mukherjee:2023qjw,Kulkarni:2023fyq,Yang:2023fhg,Zhang:2023jcf,Gomez:2023syh,Jia:2023xpx,Choudhury:2023lbp,Cao:2023juc}.
}%
, the DM relic abundance~\cite{Planck}
(either as an upper limit~\cite{CHS2,CHS4} or as a direct
measurement~\cite{CHS1,CHS3,CHS4}), 
the DM direct detection (DD) experiments~\cite{XENON,LUX,PANDAX} and 
the direct searches at the LHC~\cite{ATLAS-SUSY,CMS-SUSY}.
In \citeres{CHS1,CHS2,CHS3,CHS4} we focused on the interplay of \gmin2,
DM and future collider searches, while in \citere{gmin2-mw} the MSSM
prediction of the mass of the
$W$~boson~\cite{Heinemeyer:2006px,Heinemeyer:2007bw,Heinemeyer:2013dia}
was analyzed.

Five different scenarios were analyzed in our previous works,
classified by the mechanism that brings the LSP relic density into agreement with the measured
values. The scenarios differ by the nature of the Next-to-LSP (NLSP), or
equivalently by the hierarchies between the mass scales
determining the neutralino, chargino and slepton masses.
The relevant mass scales that determine such hierarchies
are the gaugino soft SUSY-breaking parameters $M_1$
and $M_2$, the Higgs mixing parameter $\mu$ and the slepton soft
SUSY-breaking parameters $\msl{L}$ and $\msl{R}$ (see \refse{sec:model}
for a detailed description). 
The five scenarios can be summarized as follows~\cite{CHS1,CHS2,CHS3,CHS4}:
\begin{itemize}
\item[(i)]
bino/wino DM with $\cha1$-coannihilation ($M_1 \lsim M_2$) :
DM relic density can be fulfilled, $m_{\rm (N)LSP} \lsim 650\, (700) \gev$;
\item[(ii)]
bino DM with $\Slpm$-coannihilation case-L ($M_1 \lsim \msl{L}$) :
DM relic density can be fulfilled, $m_{\rm (N)LSP} \lsim 650\, (700) \gev$;
\item[(iii)]
bino DM with $\Slpm$-coannihilation case-R ($M_1 \lsim \msl{R}$) :
DM relic density can be fulfilled, $m_{\rm (N)LSP} \lsim 650\, (700) \gev$.
\item[(iv)]
higgsino DM ($\mu < M_1, M_2, \msl{L}, \msl{R}$) :
DM relic density is only an upper bound (the full relic density implies
$\mneu1 \sim 1 \tev$ and \gmin2\ cannot be fulfilled), 
$m_{\rm (N)LSP} \lsim 500 \gev$ with $m_{\rm NLSP} - m_{\rm LSP} \sim 5 \gev$.
\item[(v)]
wino DM ($M_2 < M_1, \mu, \msl{L}, \msl{R}$) :
DM relic density is only an upper bound, (the full relic density implies
$\mneu1 \sim 3 \tev$ and \gmin2\ cannot be fulfilled),
$m_{\rm (N)LSP} \lsim 600 \gev$ with $m_{\rm NLSP} - m_{\rm LSP} \sim 0.3 \gev$.%
\footnote{It should be noted that this scenario requires special care
concerning the correct choice of the chargino/neutralino renormalization
scheme, see \citeres{Heinemeyer:2023pcc,Heinemeyer:2022apt}
and the corresponding discussion in \citere{CHS2}.}%
\end{itemize}
In all the scenarios mentioned above, a degeneracy between
the three generations of sleptons was assumed. 
Apart from the resulting simplicity of the analysis, such a degeneracy
is also motivated by the solution to the SUSY flavour problem, which
prefers the three generations of scalars to possess degenerate
soft SUSY-breaking masses~\cite{Ellis:1981ts,Barbieri:1981gn}.
Within the scenarios~(ii) 
and~(iii) the assumed degeneracy directly links the smuons
involved in the explanation  
for \gmin2\ to the  coannihilating NLSPs required to explain
the DM content of the universe.
On the other hand, in 
well-motivated top-down models with a specific high-scale
SUSY-breaking mechanism (e.g.\ minimal supergravity
(mSUGRA)~\cite{Chamseddine:1982jx}),
there is no such degeneracy, and often the 
lighter stau becomes the NLSP at the EW scale with the smuons (and
selectrons) somewhat heavier.
This has been explored in the context of the Constrained MSSM
(CMSSM) in, e.g.,
\citeres{Ellis:2003cw,Baer:2003wx,Lahanas:2003yz,Chattopadhyay:2003xi}, 
and within the non-universal scalar mass models (NUHM) in, e.g.,
\citere{Ellis:2002wv}.
Consequently, a full mass degeneracy of the three slepton
families should be regarded as an artificial constraint. 
Indeed, in \citere{Bagnaschi:2017tru} it was shown that leaving
the third generation slepton masses independent of the first and
second generations, as it had been done previously
in \citere{deVries:2015hva}, has a strong impact on the resulting
phenomenology. Therefore, it is crucial to investigate this scenario
in a general manner in the context of the EW MSSM in relation to
updated collider, DM and \gmin2 constraints.

\smallskip
In this paper we analyze an MSSM scenario at the EW scale, assuming non-universality of
the input slepton masses such that the smuons and selectrons remain mass-degenerate,
whereas the staus turn out to be lighter.
This makes the staus to be the NLSPs, enforcing stau coannihilation.
As in the scenarios~(ii) and~(iii) we either require left- or the
right-handed stau mass parameter to be close to $M_1$. For these two
scenarios, corresponding more to a 
top-down model motivated mass hierarchy, we analyze
the complementarity of DD experiments 
and future collider experiments, concretely the HL-LHC and
a possible future linear $e^+e^-$ collider, the
International Linear Collider (ILC) operated at a center-of-mass
energy of up to $\sqrt{s} \lsim 1 \tev$, the ILC1000.
In the first step we analyze the predictions for the DM relic density as a
function of the (N)LSP masses. We 
show the results both for DM fulfilling the relic density as well as
taking the DM density only as an upper bound.
In the second step we evaluate the prospects for future DD experiments
in these two scenarios. We show that both scenarios can result in DD cross
sections below the neutrino floor for a significant amount of model
parameter space with the DM relic density remaining substantially below 
the Planck measurement. In this case direct searches at the HL-LHC and
particularly at the ILC1000 will be necessary to
fully probe these scenarios. 

The plan of the paper is as follows. In \refse{sec:model} we briefly review
the parameters of the EW sector of MSSM. The relevant constraints for this
analysis are outlined in \refse{sec:constraints}. \refse{sec:paraana} contains the details
of our parameter scan strategy and analysis flow. Our results are described in \refse{sec:results}.
Finally, we summarize in \refse{sec:conclusion}.

\section {The electroweak sector of the MSSM}
\label{sec:model}

In our MSSM notation we follow exactly \citere{CHS1}, with the
exception of the degeneracy of the slepton mass parameters. We
restrict ourselves here to a very short introduction of the relevant
symbols and parameters, concentrating on 
the EW sector of the MSSM. This sector consists of charginos,
neutralinos and scalar leptons.
Concerning the scalar quark sector, we simply assume it to
be heavy such that it does not play a relevant role in our analysis.
Furthermore, throughout this 
paper we also assume the absence of $\CP$-violation, i.e.\ 
that all parameters are real.

The masses and mixings of the four neutralinos are given (on top of SM
parameters) by the $SU(2)_L$ and $U(1)_Y$
gaugino masses, $M_2$ and $M_1$, the Higgs mixing parameter $\mu$, 
as well as $\tb := v_2/v_1$: the ratio of the two vacuum expectation
values (vevs) of the two Higgs doublets.
After diagonalizing  the mass matrix the four eigenvalues
yield the four neutralino masses $\mneu1 < \mneu2 < \mneu3 <\mneu4$.
Similarly, the masses and mixings of the charginos are given (on top of SM
parameters) by $M_2$, $\mu$ and $\tb$. 
Diagonalizing the mass matrix yields the two chargino-mass
eigenvalues $\mcha1 < \mcha2$.

For the sleptons, contrary to \citeres{CHS1,CHS2,CHS3,CHS4},
we have chosen common soft SUSY-breaking parameters for the first two
generations, but different for the third generation. 
The charged selectron and smuon mass matrices are given (on top of SM
parameters) by 
the diagonal soft SUSY-breaking parameters $\mL^2$ and $\mR^2$ and the
trilinear Higgs-slepton coupling $A_l$ ($l = e, \mu$), where the
latter are set to zero. Correspondingly, the stau mass matrix is given
in terms of $\mstau{L}$, $\mstau{R}$ and $\Atau$. The latter
is scanned over a range of values determined by the vacuum stability
constraints, $\Atau^2 +3\mu^2 < 7.5 (\mstau{L}^2 + \mstau{R}^2)$,
allowing for the possibility of a metastable
universe~\cite{Gunion:1987qv,Casas:1995pd,Chattopadhyay:2014gfa,Hollik:2018wrr}.

The mixing between the ``left-handed'' and
``right-handed'' sleptons is only relevant for staus, where the
off-diagonal entry in the mass matrix is given by $-m_\tau \mu \tb$.
Consequently, for the first two generations, the mass eigenvalues can
be approximated as $\msl1 \simeq \mL, \msl2 \simeq \mR$ (assuming
small $D$-terms).
We do not mass order the sleptons, i.e.\ we follow the convention that
$\Sl_1$ ($\Sl_2$) has the large ``left-handed'' (``right-handed'')
component. As symbols for the first and second generation masses we use
$\msl1$ and $\msl2$.  We also use symbols for the scalar electron, muon
and tau masses individually, 
$\mse{1,2}$, $\msmu{1,2}$ and $\mstau{1,2}$.
The sneutrino and slepton masses are connected by the usual SU(2) relation.

Overall, the EW sector at the tree level
can be described with the help of nine parameters: $M_2$, $M_1$, $\mu$,
$\tb$, $\mL$, $\mR$, $\mstau{L}$, $\mstau{R}$ and $\Atau$.
We assume $\mu, M_1, M_2 > 0 $ throughout our analysis.
In \citere{CHS1} it was shown that this covers the relevant parameter space
once the \gmin2\ results are taken into account
(see, however, the discussion in \citere{Baum:2021qzx}).


\medskip
Following the experimental limits for strongly interacting particles from the
LHC~\cite{ATLAS-SUSY,CMS-SUSY}, 
we assume that the colored sector of the MSSM is substantially heavier
than the EW sector, and therefore does not play a role in our
analysis. For the Higgs-boson sector we assume that the radiative
corrections to the light 
$\cp$-even Higgs boson, originating largely from the top/stop
sector, yield a value in agreement with the experimental data,
$\Mh \sim 125 \gev$. This naturally yields stop masses in the TeV
range~\cite{Bagnaschi:2017tru,Slavich:2020zjv}, in agreement 
with the LHC bounds. Concerning the heavy Higgs-boson mass scale, 
$\MA$ (the $\cp$-odd Higgs-boson mass), we have shown
in \citeres{CHS1,CHS2,CHS3,CHS4} that $A$-pole annihilation is largely
excluded. Consequently, we also here we assume $\MA$ to be sufficiently
large to not play a role in our analysis.


\section {Relevant constraints}
\label{sec:constraints}

The SM prediction of \amu\ is given by~\cite{Aoyama:2020ynm}
(based on \citeres{Aoyama:2012wk,Aoyama:2019ryr,Czarnecki:2002nt,Gnendiger:2013pva,Davier:2017zfy,Keshavarzi:2018mgv,Colangelo:2018mtw,Hoferichter:2019mqg,Davier:2019can,Keshavarzi:2019abf,Kurz:2014wya,Melnikov:2003xd,Masjuan:2017tvw,Colangelo:2017fiz,Hoferichter:2018kwz,Gerardin:2019vio,Bijnens:2019ghy,Colangelo:2019uex,Blum:2019ugy,Colangelo:2014qya}),
\begin{align}
\amu^{\rm SM} &= (11 659 181.0 \pm 4.3) \times 10^{-10}~.
\label{gmt-sm}
\end{align}
After the publication of the last results from the 
Muon g-2 Theory Initiative~\cite{Aoyama:2020ynm}, a lattice calculation~\cite{Borsanyi:2020mff} 
for the leading order hadronic vacuum polarization (LO HVP) contribution has appeared
yielding a somewhat higher value for the $\amu^{\rm SM}$.
While this result is partially supported by some other lattice
calculations~\cite{Ce:2022kxy,ExtendedTwistedMass:2022jpw}, a consensus among the various lattice groups
is yet to be established. On the other hand, no new SM theory prediction has been
published so far. In this analysis, we do
not take the lattice result into account, and use \refeq{gmt-sm} for the theoretical SM prediction of \amu~
(see also the discussions in \citeres{CHS1,Lehner:2020crt,Borsanyi:2020mff,Crivellin:2020zul,Keshavarzi:2020bfy,deRafael:2020uif}).%
\footnote{
On the other hand, it is obvious that our conclusions would change
substantially if the result presented in \cite{Borsanyi:2020mff}
turned out to be correct.}%

The combined experimental world average, based
on \citeres{Abi:2021gix,Bennett:2006fi}, is given by
\begin{align}
\amu^{\rm exp} &= (11 659 206.1 \pm 4.1) \times 10^{-10}~.
\label{gmt-exp}
\end{align}
Compared with the SM prediction in \refeq{gmt-sm}, one finds a 
deviation of
\begin{align}
\Delta\amu &= (25.1 \pm 5.9) \times 10^{-10}~, 
\label{gmt-diff}
\end{align}
corresponding to a $4.2\,\sig$ discrepancy.
We use this limit as a cut at the $\pm2\,\sig$ level.

\medskip
In the MSSM the main contribution to \gmin2\ originates from one-loop
diagrams involving $\neu1-\tilde \mu$ and $\cha1-\Sn$
loops~\cite{Lopez:1993vi,Chattopadhyay:1995ae,Chattopadhyay:2000ws,Kowalska:2015zja}. 
In our analysis the MSSM contribution to \gmin2\ is based on a full
one-loop plus partial two-loop
calculation~\cite{vonWeitershausen:2010zr,Fargnoli:2013zia,Bach:2015doa}
(see also \cite{Heinemeyer:2003dq,Heinemeyer:2004yq}), as implemented
into the code {\tt GM2Calc}~\cite{Athron:2015rva}.

\bigskip
All other constraints are taken into account exactly as
in \citere{CHS1,CHS2,CHS3,CHS4}. These are :

\begin{itemize}

\item Vacuum stability constraints:\\
Our parameter scan ranges (see \refeqs{slep-coann-doublet} and (\ref{slep-coann-singlet}))
are determined keeping the vacuum stability constraint
in mind. On top of that, all points are checked to possess a correct and stable EW vacuum, e.g.\
avoiding charge and color breaking minima, employing
the public code {\tt Evade}~\cite{Hollik:2018wrr,Ferreira:2019iqb}.

\item Constraints from the LHC:\\
All relevant SUSY searches for EW particles are taken into account,
mostly via \CM~\cite{Drees:2013wra,Kim:2015wza,Dercks:2016npn} (see 
\citere{CHS1} for details on many analyses newly implemented by our group).
In the following we briefly review the relevance of various LHC searches
for the present analysis.
\begin{itemize}
\item
The production of $\cha1-\neu2$ pairs leading to three leptons and
$\met$ in the final state~\cite{ATLAS:2019wgx,ATLAS:2021moa}.
We have implemented in \texttt{CheckMATE} the chargino-neutralino
pair production searches in final states with three leptons and
missing transverse momentum from \citere{ATLAS:2021moa}.
We have included only the on-shell $WZ$
selection which is the most important mode for our analysis.

\item
Slepton-pair production leading to
two same flavour opposite sign leptons and $\met$ in the final state~\cite{Aad:2019vnb}.

\item
The ATLAS and CMS searches for direct stau pair production target
the mass gap region $\mstau{L,R} -\mneu1 \gtrsim 100 \gev$~\cite{ATLAS:2019gti,CMS:2022rqk,ATLAS:2023djh}.
This bound may, in principle, be relevant for the pair production of the heavier staus i.e.\
$\tilde \tau_R$ ($\tilde \tau_L$) for stau-L (stau-R) case (see \refse{sec:results}).  
However, we have explicitly checked that after the application of \gmin2\,, DM and
LHC constraints mentioned above, the surviving parameter points in our scans stay
beyond the sensitivity reach of the stau pair production seraches.

The pair production of $\cha1-\neu2$ pairs decaying via staus~\cite{ATLAS:2023djh}
may be effective in constraining our parameter space. However, the decay products
from the lighter stau, the coannihilation partner of the $\neu1$,
will be too soft to be detected by these searches. Thus, only the heavier one
of the two staus may contribute to the signal cross section, provided the decay of $\cha1-\neu2$
via stau is kinematically allowed.
Filtering out the parameter points for which the heavier stau is lighter than  $\cha1$ and $\neu2$,
we observed that only a handful points in the low $\cha1$/$\neu2$ mass region
may potentially be affected by this constraint. A detailed account of the impact
of the stau searches~\cite{ATLAS:2023djh} on our parameter space is reserved for a future analysis.

The latest bounds from the compressed stau searches are far too weak at present to be
of relevance for our analysis~\cite{CMS:2019zmn}.

\item
The low mass gap between the third generation sleptons and the lightest
neutralino in our analysis may give rise to long lived staus which
are subject to bounds from dedicated long-lived particle (LLP) searches at the LHC.
The ATLAS collaboration has looked for a heavy stable charged particle (HSCP) through specific ionisation
energy loss in the detector~\cite{Heinrich:2018pkj,ATLAS:2019gqq}.
The results have been interpreted for the pair
production of staus in a gauge mediated SUSY-breaking scenario (GMSB) assuming
stable staus (see \refse{sec:staulifetime} for detail).
Since the search strategy is largely independent of the undelying model assumption,
this limit can be applied to constrain the long lived staus in our model.
The disappearing track searches~\cite{ATLAS:2022rme, CMS:2023czm} are
targeted towards long-lived  winos or higgsinos with production
cross-section much larger compared to  the staus. Thus, this search is ineffective in constraining our model
parameter space. Our scenario remains equally unaffected by the displaced lepton
searches~\cite{ATLAS:2020wjh,CMS:2022rqk} which requires a high-$p_T$ displaced lepton
in the signal events, following the theoretical framework of GMSB scenarios.

\end{itemize}

\item
Dark matter relic density constraints:\\
the latest result from Planck~\cite{Planck} provides the experimental
data. The relic density is given as
\begin{align}
\Omega_{\rm CDM} h^2 \; = \; 0.120  \pm 0.001 \, , 
\label{OmegaCDM}
\end{align}
which we use as a measurement of the full MSSM density, 
or as an upper bound (evaluated with the central value
plus $2\sigma$), 
\begin{align}
\Omega_{\rm CDM} h^2 \; \le \; 0.122 \, . 
\label{OmegaCDMlim}
\end{align}
The evaluation of the relic density in the MSSM is performed with
\MO~\cite{Belanger:2001fz,Belanger:2006is,Belanger:2007zz,Belanger:2013oya}.

\item
Direct detection constraints of Dark matter:\\
In comparison to previous analyses we use an updated limit on 
the spin-independent (SI)
DM scattering cross-section $\ssi$ from the LZ~\cite{LZ-new}
experiment (which are always substantially more
relevant than the spin-dependent limits). 
The theoretical predictions are evaluated using the public code
\MO~\cite{Belanger:2001fz,Belanger:2006is,Belanger:2007zz,Belanger:2013oya}.
Apart from this limit we will discuss the impact of possible
future limits and the neutrino floor below.

For parameter points with $\Omega_{\tilde \chi} h^2 \; \le \; 0.118$
(i.e.\  lower than the $2\,\sig$ lower limit from Planck~\cite{Planck}, 
see \refeq{OmegaCDMlim})
the DM scattering cross-section is rescaled
with a factor of ($\Omega_{\tilde \chi} h^2$/0.118). 
This takes into account the fact that $\neu1$ provides only a fraction of the
total DM relic density of the universe.

\end{itemize}

Another potential set of constraints is given by the indirect
detection of DM. However, we do not impose these constraints
on our parameter space because of the well-known large uncertainties
associated with astrophysical factors like DM density profile as
well as theoretical corrections,
see~\citeres{Slatyer:2017sev,Hryczuk:2019nql,Rinchiuso:2020skh,Co:2021ion}. 
The most precise indirect detection limits come from DM-rich
dwarf spheroidal galaxies, where the uncertainties on the cross section
limits are found in the
range of $\sim 2-3$ (from the lowest to the highest cross section
limit)~\cite{McDaniel:2023bju,Fermi-LAT:2015att}. The 
most recent analysis~\cite{McDaniel:2023bju}, assuming the
annihiliation goes into one single mode (which is too
``optimistic'' for our scenarios), sets limits of
$\mneu1 \gsim 100 \gev$ for the generic thermal relic
saturating \refeq{OmegaCDM}. However, this is
well below our preferred parameter space (see next sections below).
Additionally, it has been noted 
previously~\cite{Chakraborti:2014gea,Baer:2016ucr}
that the indirect detection cross section of the DM for the stau
coannihilation scenario lies almost two orders of magnitude below the
current limits from Fermi-LAT~\cite{McDaniel:2023bju}.
Consequently, we do not further consider the indirect
detection constraints in our analysis.


\section{Parameter scan and analysis flow}
\label{sec:paraana}

\subsection{Parameter scan}
\label{sec:scan}

We scan the EW MSSM parameter space, fully covering the allowed
regions of the relevant neutralino, chargino, stau  and first/second
generation slepton masses.
We follow the approach taken in \citeres{CHS1,CHS2,CHS3,CHS4} and
investigate the two scenarios of stau coannihilation discussed
in \refse{sec:intro}. They are 
given by the possible mass orderings of $M_1 < M_2, \mu$, and $\mstau{L}$,
$\mstau{R}$. These masses yield a bino-like LSP and fix the NLSP,
thus ensuring stau coannihilation as the mechanism that reduces the relic
DM density in the early 
universe to or below the current value (see \refeqs{OmegaCDM},
(\ref{OmegaCDMlim})). We do not consider 
the possibility of pole annihilation, e.g.\ with the
$h$, the $A$ or the $Z$~boson. As argued in \citeres{CHS1,CHS2,CHS3,CHS4}
these are rather remote possibilities in our set-up.%
\footnote{
Concretely, we have set $\MA = 1.5 \tev$, ensuring that the heavy
Higgs-boson sector does not play a role in our analysis.
}%

As indicated above, we choose $M_1$ to be the smallest mass
parameter and require that a scalar tau is close in mass.
In this scenario ``accidentally'' the wino or higgsino
component of the $\neu1$ can be non-negligible in some parts of the parameter space.
However, this is not a distinctive feature of this scenario.
We distinguish two cases: either the SU(2)
doublet staus, or the singlet staus are close in mass to the LSP.\\

\noindent
{\bf stau-L: bino DM with \boldmath{$\Stau1$}-coannihilation} (SU(2) doublet)
\begin{align}
  100 \gev \leq M_1 \leq 1000 \gev \;,
  \quad 1.2 \, M_1 \leq M_2 \leq 10 \, M_1 \;, \notag\\
  \quad 1.2 \, M_1 \leq \mu \leq 10 \, M_1, \;
  \quad 5 \leq \tb \leq 60 \;, \notag\\
  \quad M_1 \leq \mstau{L} \leq 1.2 \, M_1, 
  \quad \mstau{L} \leq \mstau{R} \leq 10 \, \mstau{L} \;, \notag\\
  \quad 1.2 \, M_1 \leq \mL,\mR \leq 2 \tev \;, \notag\\
  \Atau^2 \leq 7.5 (\mstau{L}^2 + \mstau{R}^2) - 3 \mu^2~.
\label{slep-coann-doublet}
\end{align}

\noindent
{\bf stau-R: bino DM with \boldmath{$\Stau2$}-coannihilation} (SU(2) singlet)
\begin{align}
  100 \gev \leq M_1 \leq 1000 \gev \;,
  \quad 1.2 M_1 \leq M_2 \leq 10 M_1 \;, \notag \\
  \quad 1.2 M_1 \leq \mu \leq 10 M_1, \;
  \quad 5 \leq \tb \leq 60 \;, \notag\\
  \quad \mstau{R} \leq \mstau{L} \leq 10 \mstau{R}, \; 
  \quad M_1 \leq \mstau{R} \leq 1.2 M_1 \;, \notag\\
  \quad 1.2 M_1 \leq \mL,\mR \leq 2 \tev\;, \notag\\
  \Atau^2 \leq 7.5 (\mstau{L}^2 + \mstau{R}^2) - 3 \mu^2~.
\label{slep-coann-singlet}
\end{align}

\noindent
In both scans we choose flat priors of the parameter space and
generate \order{10^7} points.
In order to obtain reliable {\it upper} limits on the (N)LSP
masses, we performed dedicated scans in the respective parameter
regions. These appear as more densely populated regions in the plots
below.

As discussed above, the mass parameters of the colored sector have
been set to high values, moving these particles 
outside the reach of the LHC.
The mass of the lightest $\cp$-even Higgs-boson is in agreement with the LHC
measurements of the $\sim 125 \gev$ (concrete values are not relevant
for our analysis). Similarly, $\MA$ has been set to be above the TeV
scale (see above).


\subsection{Analysis flow}
\label{sec:flow}

The two data samples were generated by scanning randomly over the input
parameter ranges given above, assuming a flat prior for all parameters.
We use the code {\tt SuSpect-v2.43}~\cite{Djouadi:2002ze,Kneur:2022vwt}
as spectrum and SLHA file generator. In this step we ensure that all points 
satisfy the $\chapm1$ and slepton mass limits from LEP~\cite{lepsusy}.
The SLHA output files as generated by
{\tt SuSpect} are then passed as input files to \MO\,-\texttt{v5.2.13} and {\tt GM2Calc-v2.1.0} for
the calculation of the DM observables and \gmin2 respectively. The parameter
points that satisfy the \gmin2\ constraint of \refeq{gmt-diff}, the DM
relic density constraint of \refeq{OmegaCDM} or (\ref{OmegaCDMlim}), 
the DD constraints (possibly with a
rescaled cross section) and the vacuum stability constraints, tested
with {\tt Evade}, are then passed to the final check 
against the LHC constraints as implemented in \CM.
The relevant branching ratios of the SUSY particles (required by \CM)
are calculated using {\tt SDECAY-v1.5a}~\cite{Muhlleitner:2003vg}.


\section{Results}
\label{sec:results}

We follow the analysis flow as described above
and denote the points surviving certain constraints
with different colors:
\begin{itemize}
\item grey (round): all scan points.
\item green (round): all points that are in agreement with \gmin2, taking
into account the limit as given in \refeq{gmt-diff}, but are excluded by
the DM relic density.
\item blue (triangle): points that additionally give the correct relic density,
see \refse{sec:constraints}, but are excluded by the DD constraints.
\item cyan (diamond): points that additionally pass the DD constraints,
see \refse{sec:constraints}, but are excluded by the LHC constraints.
\item red (star):  points that additionally pass the LHC constraints,
see \refse{sec:constraints}. 
\end{itemize}


\subsection{Upper limits and preferred parameter ranges :  Stau-L case}
\label{sec:stauL}

We start our phenomenological analysis with the case of bino DM with
$\Stau1$-coannihilation. In this scenario the $\mstau{L}$ parameter is
close to $M_1$, defining the NLSP and the coannihilation mechanism.
We refer to this scenario as the stau-L case.

\begin{figure}[htb!]
       \vspace{1em}
\centering
\begin{subfigure}[b]{0.48\linewidth}
\centering\includegraphics[width=\textwidth]{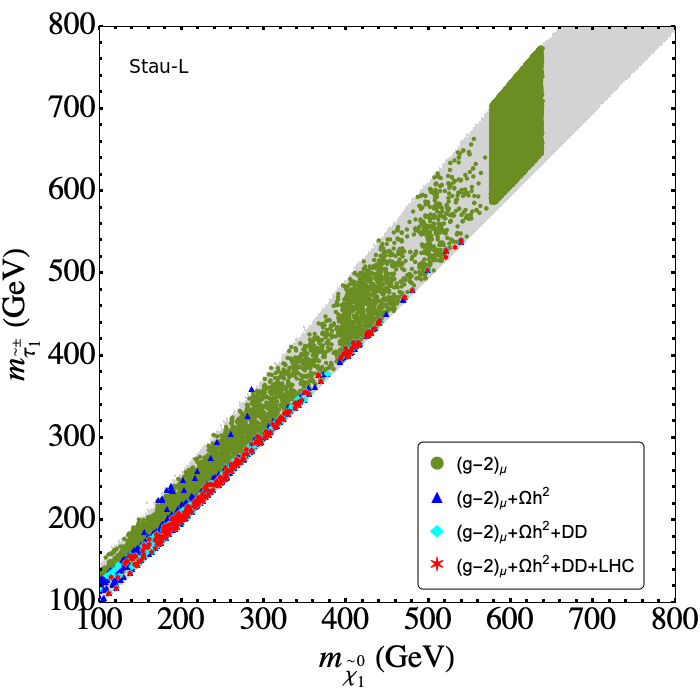}
        \caption{}
        \label{}
\end{subfigure}
~
\begin{subfigure}[b]{0.48\linewidth}
\centering\includegraphics[width=\textwidth]{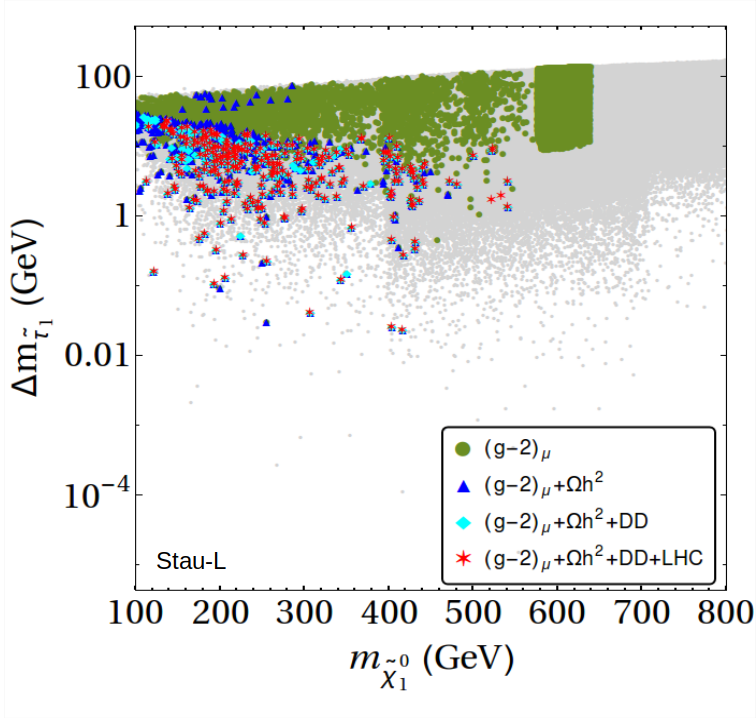}
        \caption{}
        \label{}
\end{subfigure}
       \vspace{1em}
\caption{The results of our parameter scan in the $\Stau1$
coannihilation case in the $\mneu1$--$m_{\Stau1}$ plane
(left) and the $\mneu1$--$\De m_{\Stau1}$ 
plane ($\De m_{\Stau1} = \mstau1 - \mneu1$, right plot).
For the color coding: see text.
}
\label{mn1-mstau1-L}
\end{figure}

\begin{figure}[htb!]
       \vspace{1em}
\centering
\begin{subfigure}[b]{0.48\linewidth}
\centering\includegraphics[width=\textwidth]{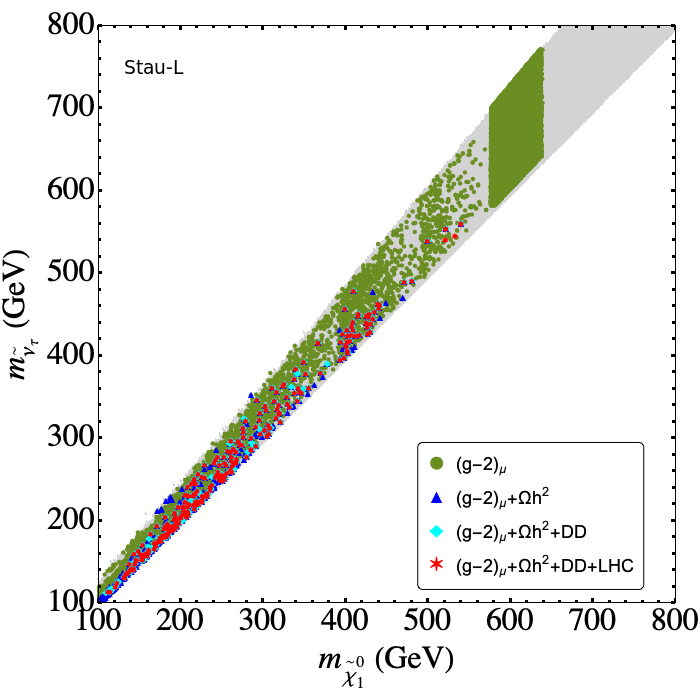}
        \caption{}
        \label{}
\end{subfigure}
~
\begin{subfigure}[b]{0.48\linewidth}
\centering\includegraphics[width=\textwidth]{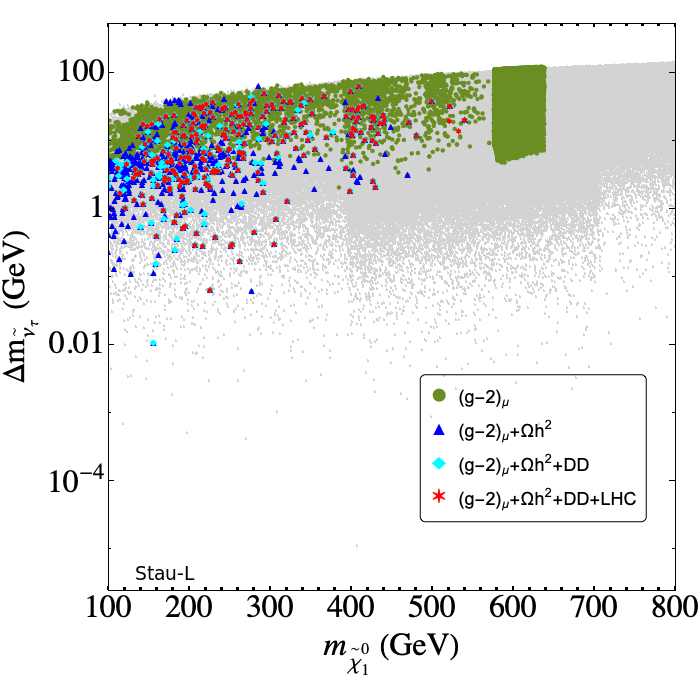}
        \caption{}
        \label{}
\end{subfigure}
       \vspace{1em}
\caption{The results of our parameter scan in the
$\Stau1$-coannihilation case in the 
$\mneu1$--$\msnutau$ plane (left) and the 
$\mneu1$--$\De m_{\sntau}$ plane ($\De m_{\sntau} = \msnutau - \mneu1$).}
\label{mn1-msneu-L}
\end{figure}

In \reffi{mn1-mstau1-L} we show the result of our parameter scan in
the $\mneu1$--$m_{\Stau1}$ plane (left) and the $\mneu1$--$\De m_{\Stau1}$ 
plane ($\De m_{\Stau1} := \mstau1 - \mneu1$, right plot). In the left
plot the points are found by definition close to the diagonal. The extra scan region
is clearly visible around $\mneu1 \sim 600 \gev$ in green,
i.e.\ in agreement with \gmin2. The fact that our scan is effectively
exhaustive w.r.t.\ \gmin2 can be understood simply by looking at the lower plot
of \reffi{mn1-other-L} (see below), where we depict our scanned points in $\mneu1-\tb$ plane.
The upper limit on $\mneu1$ from \gmin2  is expected to be reached for
the largest value of $\tb$ in our scans i.e. $\tb \approx 60$. By looking at the
general behaviour of the points it can be inferred that the points can go at the most up to
$\mneu1 \approx 660 \gev$, saturating $\tb \approx 60$.
Thus, the $\mneu1$ range not covered in the scans amounts to $\sim 20 \gev$.
Furthermore, since the final surviving points (red) indicate a clear upper limit,  we refrained from scanning for even higher
$\mneu1$ values. The most restrictive constraint after the application of \gmin2\ comes in the next step,
i.e. with the requirement of correct relic density (dark blue+cyan+red points).
This sets an upper limit on $\mneu1$ and $\mstau1$
of $\sim 550 \gev$. As can be seen in the right plot, the relic density
constraint also requires a small mass splitting, which is decreasing
with increasing $\mneu1$. This is similar to the ``stau coannihilation
strip'' discussed in \citere{Ellis:2001msa}.
The only exception are a few points for $\mneu1 \lsim 300 \gev$ where the mass difference
can be relatively larger. These points correspond to a substantial bino-higgsino and/or bino-wino
mixing, making stau coannihilation only a subdominant component of
the total annihilation cross-section.
As we go towards higher $\mneu1$, $\tan\beta$  becomes restricted
to larger values, determining the mass difference between $\neu1$ the $\Stau1$.
The application of DD constraints (cyan+red points), 
does not have an impact on the allowed parameter space in the
$\mneu1$-$\mstau1$ plane, except for cutting away the points $\mneu1 \lsim 300 \gev$
with larger mass differences. For these points, the proximity of $M_1$, $\mu$ and/or $M_2$ makes
the $\ssi$ sufficiently large~\cite{Hisano:2004pv} so that they are excluded by the LZ~\cite{LZ-new} bounds.
Finally, the LHC constraints cut away mainly
points with low $\mneu1$, resulting in the red points which are in
agreement with all available constraints. Overall, we find
$m_{\rm (N)LSP} \lsim 550 \gev$ and $\De\mstau1 \lsim 30 \gev$.

The results look very similar in the $\mneu1$-$\msnutau$ plane and
$\mneu1$-$\De \msnutau$ plane ($\De\msnutau := \msnutau - \mneu1$), as
show in the left and right plots of \reffi{mn1-msneu-L} respectively.
Compared to \reffi{mn1-mstau1-L},
here somewhat larger mass differences are found for the smaller $\mneu1$ values. However, the
upper bound found is similar as in \reffi{mn1-mstau1-L},
$\msnutau \lsim 550 \gev$.

Next, in \reffi{mn1-other-L} we show the results of our parameter scan in the
$\mneu1$--$\msmu1$ plane (top left), the $\mneu1$--$\mcha1$ plane (top
right) and the $\mneu1$--$\tb$ plane (bottom). Since we enforce
$\Stau1$-coannihilation, the other masses are effectively free. However,
the combination of \gmin2\ and LHC constraints still yield an upper
limit on the lighter smuon mass of $\msmu1 \lsim 800 \gev$, i.e.\ only
about $\sim 100 \gev$ higher than in the case of
$\Smu1$-coannihilation~\cite{CHS3}. On the other hand, the lighter chargino mass driven by $M_2$ and $\mu$, 
corresponds to hardly any upper limit and values exceeding $\sim 3 \tev$ are found. 

We finish our analysis of the $\Stau1$-coannihilation case with the
$\mneu1$-$\tb$ plane presented in the lower plot of \reffi{mn1-other-L}.
The \gmin2\ constraint is
fulfilled in a triangular region with largest neutralino masses allowed
for the largest $\tb$ values (where we stopped our scan at $\tb = 60$).
In agreement with the previous plots, the largest values for the
lightest neutralino masses allowed by all the constraints are $\sim 550 \gev$. 
The LHC constraints cut out points at low $\mneu1$, nearly
independent of $\tb$, but disallowing all points with $\tb \lsim 15$.
One can observe that points with masses of
$\mneu1 \sim \mstau2 \sim 120 \gev$ are still allowed by the LHC searches.
As mentioned in \refse{sec:constraints}, the current sensitivity of the
compressed stau searches~\cite{CMS:2019zmn} are not sufficiently high to provide any
constraint on our parameter space. Although slepton pair production searches~\cite{Aad:2019vnb}
are able to provide some constraints,
the limits from the $\cha1-\neu2$ pair production searches~\cite{ATLAS:2019wgx,ATLAS:2021moa}
gets relaxed because of the substantially large branching fraction of the $\cha1$ and $\neu2$ via the staus.
In this plot we also show as a black line the bound
from the latest ATLAS search for heavy neutral Higgs
bosons~\cite{Aad:2020zxo} in the channel $pp \to H/A \to \tau\tau$
in the $M_h^{125}(\tilde\chi)$ benchmark scenario~\cite{Bagnaschi:2018ofa}%
\footnote{This bound is based on the full Run~2 data from ATLAS, 
using the version~5 of
{\tt HiggsBounds}~\cite{Bechtle:2008jh,Bechtle:2011sb,Bechtle:2013wla,Bechtle:2015pma,Bechtle:2020pkv,Bahl:2022igd}.
Subsequently, a corresponding limit from CMS was
published~\cite{CMS:2022goy}, which is somewhat stronger in particlular
for lower Higgs mass values, strengthening our argument.}%
. The black line in the plane has been obtained setting $\mneu1 = \MA/2$,
i.e.\ roughly to the requirement for $A$-pole
annihilation, where points above the black lines are experimentally excluded. 
There are no points passing the current \gmin2\ constraint
below the black $A$-pole line. Consequently, $A$-pole annihilation can
be considered as excluded in this scenario.

\begin{figure}[htb!]
       \vspace{4em}
\centering
\begin{subfigure}[b]{0.48\linewidth}
\centering\includegraphics[width=\textwidth]{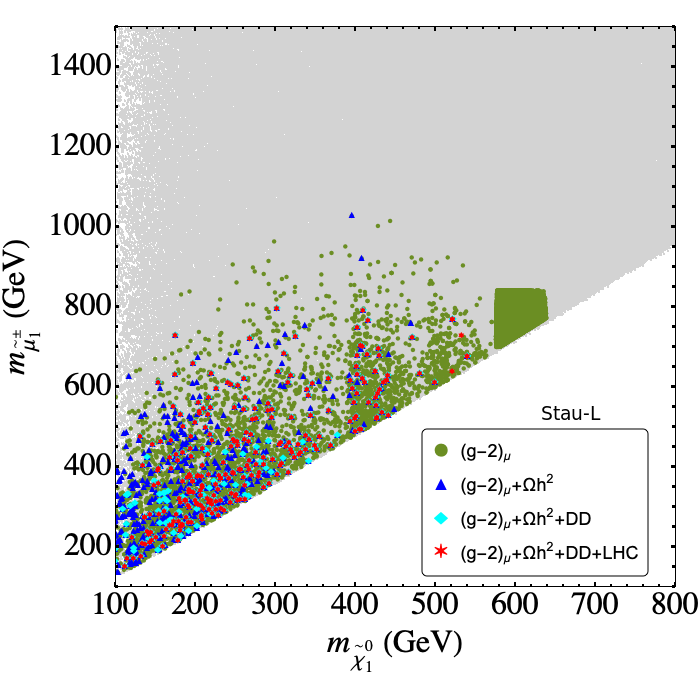}
        \caption{}
        \label{}
\end{subfigure}
~
\begin{subfigure}[b]{0.48\linewidth}
\centering\includegraphics[width=\textwidth]{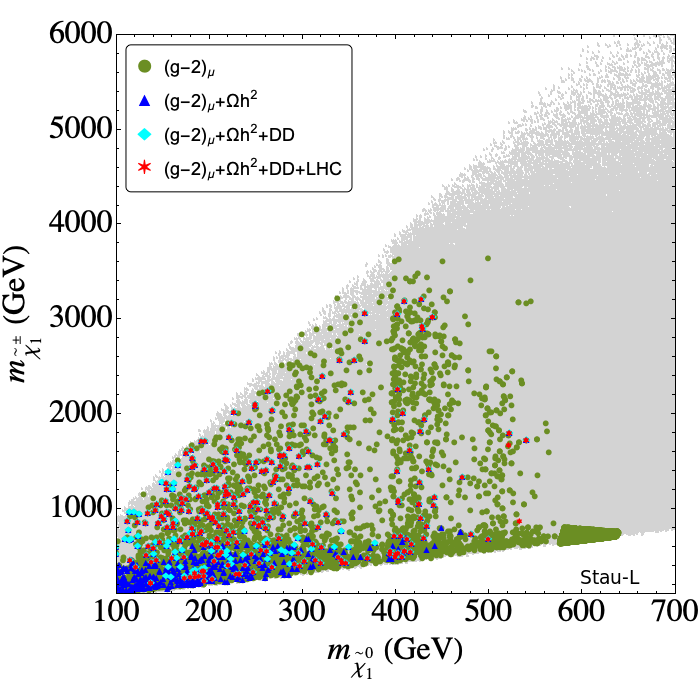}
        \caption{}
        \label{}
\end{subfigure}

\vspace{3em}
\begin{subfigure}[b]{0.48\linewidth}
\centering\includegraphics[width=\textwidth]{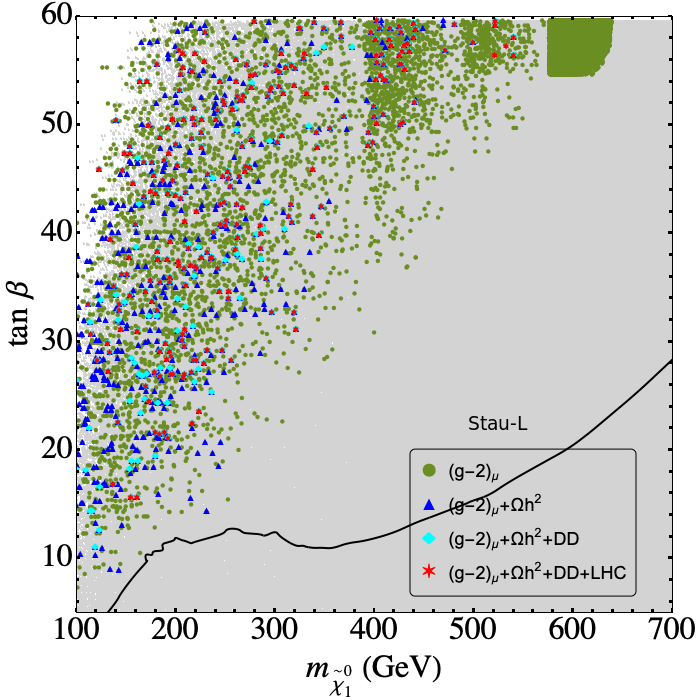}
        \caption{}
        \label{}
\end{subfigure}
\vspace{1em}
\caption{The results of our parameter scan in the
        $\Stau1$-coannihilation case in the
        $\mneu1$--$\msmu1$ plane (top left),
        $\mneu1$--$\mcha1$ plane (top right) and
        $\mneu1$--$\tb$ plane (bottom).
For the color coding: see text.
}
\label{mn1-other-L}
\end{figure}


\subsection{Upper limits and preferred parameter ranges : Stau-R case}
\label{sec:stauR}

The second part of our phenomenological analysis is the case of bino DM with
$\Stau2$-coannihilation. In this scenario, referred to as the stau-R case,
the $\mstau{R}$ parameter is
close to $M_1$, defining the NLSP and the coannihilation mechanism.

\begin{figure}[htb!]
       \vspace{1em}
\centering
\begin{subfigure}[b]{0.48\linewidth}
\centering\includegraphics[width=\textwidth]{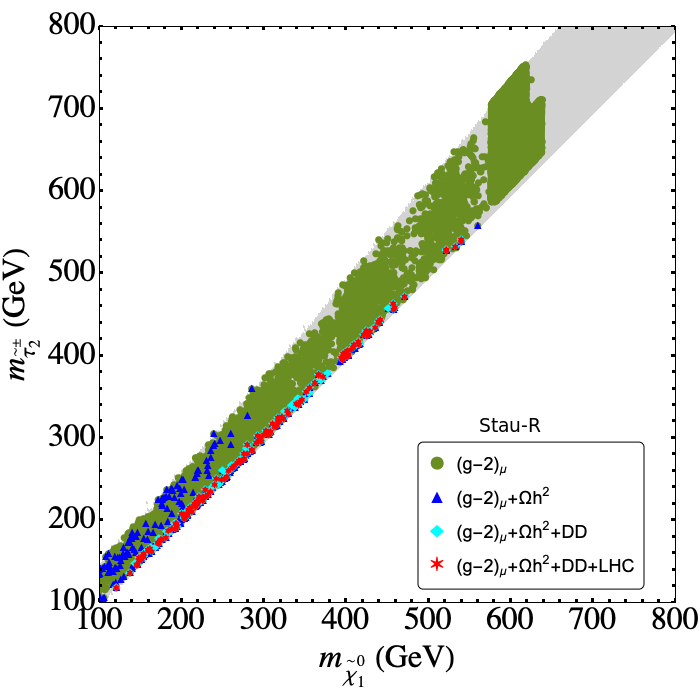}
        \caption{}
        \label{}
\end{subfigure}
~
\begin{subfigure}[b]{0.48\linewidth}
\centering\includegraphics[width=\textwidth]{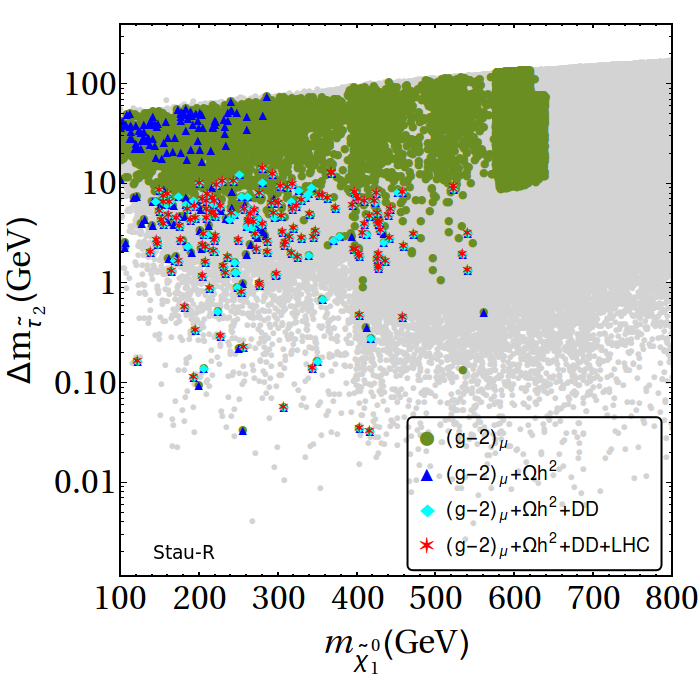}
        \caption{}
        \label{}
\end{subfigure}
\caption{
The results of our parameter scan in the $\Stau2$-coannihilation
case in the $\mneu1$--$m_{\Stau2}$ plane
(left) and the $\mneu1$--$\De m_{\Stau2}$
plane ($\De m_{\Stau2} = \mstau2 - \mneu1$, right plot).
For the color coding: see text.}
\label{mn1-mstau2-msntau-R}
\end{figure}

In \reffi{mn1-mstau2-msntau-R} we show the result of our parameter scan in
the $\mneu1$--$m_{\Stau2}$ plane (left) and the $\mneu1$--$\De \mstau2$ 
plane ($\De \mstau2 := \mstau2 - \mneu1$, right plot). In the left
plot, as in the stau-L case,
the points are found by definition close to the diagonal. Once again, one can clearly see
the densely scanned region around $\mneu1 \sim 600 \gev$ in green,
i.e.\ in agreement with \gmin2. The same conclusions as for the stau-L case holds also in this case.
The requirement of correct relic density sets an upper limit
on $\mneu1$ and $\mstau2$ of $\sim 550 \gev$. As can be seen in the right plot, the relic density
constraint requires a small mass splitting, $\De\mstau2 \lsim 10 \gev$. As in the stau-L case, a
few atypical points with larger mass differences are found for $\mneu1 \lsim 300 \gev$, where
the mixing among bino, wino and higgsino can be relatively large.
This results in their exclusion by the DD constraints (cyan+red).
Finally, the LHC constraints have hardly any impact on the allowed
parameter space. Even points with masses of
$\mneu1 \sim \mstau2 \sim 120 \gev$ are permitted by the latest LHC constraints for reasons similar to
those described in the ontext of the stau-L scenario.
Overall, we find
$m_{\rm (N)LSP} \lsim 550 \gev$ and $\De\mstau2 \lsim 10 \gev$.

\begin{figure}[htb!]
       \vspace{3em}
\centering
\begin{subfigure}[b]{0.48\linewidth}
\centering\includegraphics[width=\textwidth]{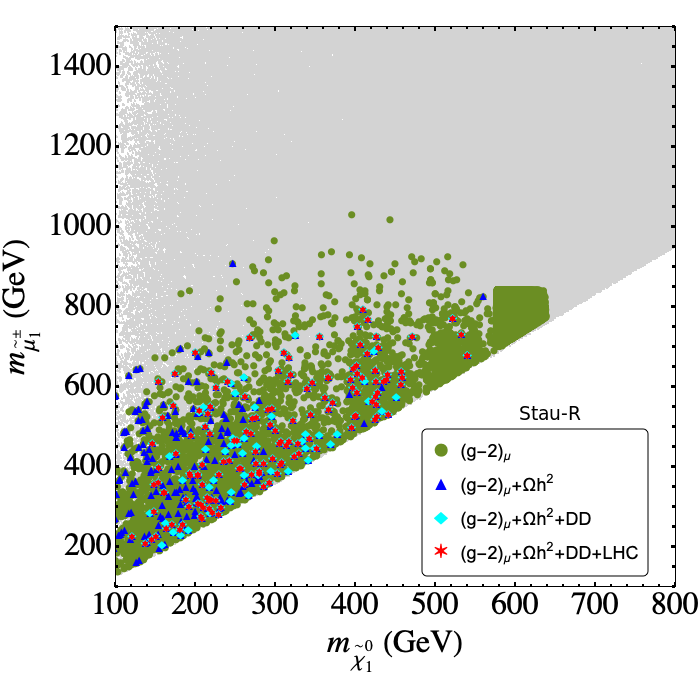}
        \caption{}
        \label{}
\end{subfigure}
~
\begin{subfigure}[b]{0.48\linewidth}
\centering\includegraphics[width=\textwidth]{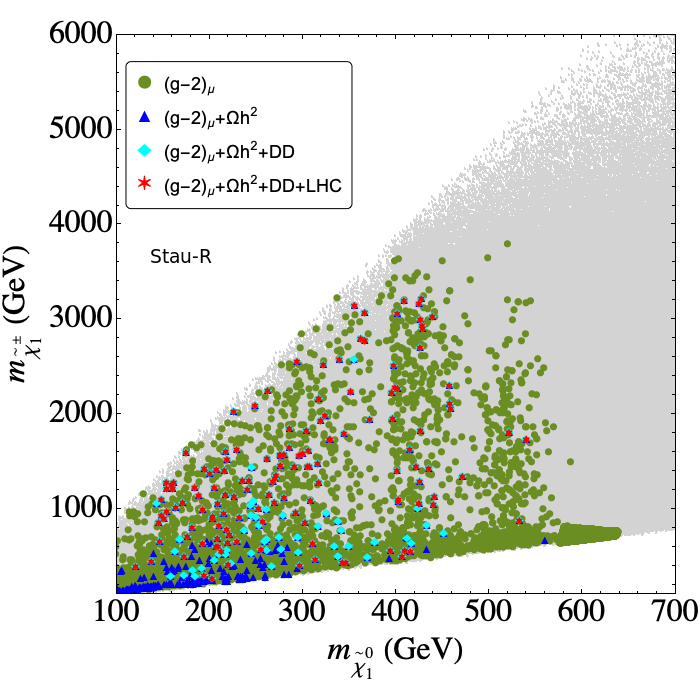}
        \caption{}
        \label{}
\end{subfigure}

\vspace{4em}
\begin{subfigure}[b]{0.48\linewidth}
\centering\includegraphics[width=\textwidth]{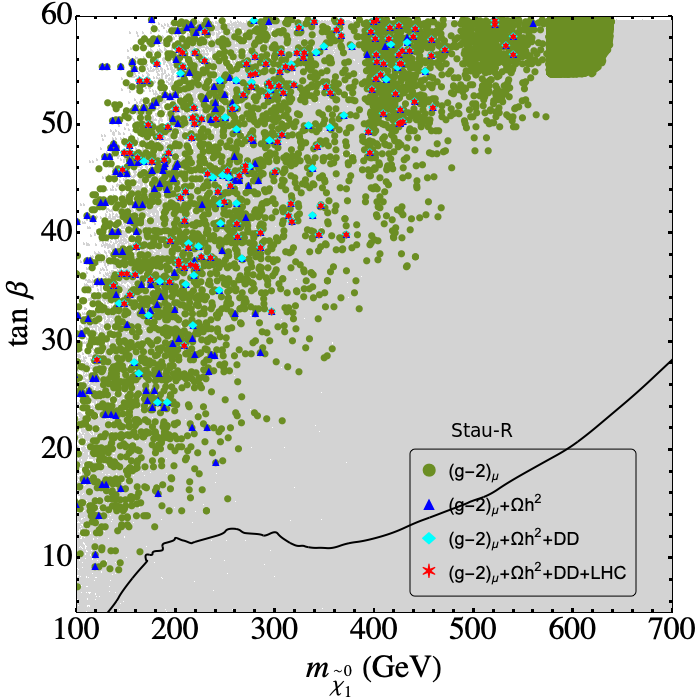}
        \caption{}
        \label{}
\end{subfigure}
\vspace{1em}
\caption{The results of our parameter scan $\Stau2$-coannihilation case
in the $\mneu1$--$\msmu1$ plane (top left),
        $\mneu1$--$\mcha1$ plane (top right) and
        $\mneu1$--$\tb$ plane (bottom).
For the color coding: see text.
}
\label{mn1-other-R}
\end{figure}

Next, in \reffi{mn1-other-R} we show the results our parameter scan in the
$\mneu1$--$\msmu1$ plane (top left), the $\mneu1$--$\mcha1$ plane (top
right) and the $\mneu1$--$\tb$ plane (bottom).
The results are very similar to the stau-L case, and we refrain here
from detailing them. Overall, we find $\msmu1 \lsim 800 \gev$, about 
about $\sim 100 \gev$ higher than in the case of
$\Smu2$-coannihilation~\cite{CHS3}. For the chargino mass we find
$\mcha1 \lsim 3 \tev$. Also the results in the $\mneu1$-$\tb$ plane are
very similar, but now we find $\tb \gsim 25$. In this case also the
$A$-pole annihilation can 
be considered as excluded in this scenario.


\subsection{Stau finite life time}
\label{sec:staulifetime}

In this section we discuss the impact of the small mass difference
between the $\neu1$ LSP and the $\Stau{}$ NLSP, which can lead to a
``finite stau life time''. 
To show the impact of the LLP and stable particle%
\footnote{``Stable'' should be seen as ``detector-stable'',
corresponding to a lifetime of larger than $\sim 100$~ns.}%
~searches at the LHC on our parameter space,
we calculate the lifetime of the staus following \citere{Jittoh:2005pq}%
\footnote{The additional decay modes included
in \citere{Citron:2012fg} are expected to have only marginal effects
in our lifetime values and no effect on our conclusion regarding 
the stable charged particle searches at the LHC.}%
. The results are shown in \reffi{lifetime-delm},
in lifetime vs.\ $\Delta m = \mstau{1,2} - \mneu1$ 
plane for $\Stau1$ (stau-L) and $\Stau2$ (stau-R).
The vertical dotted line indicates $\Delta m = \mstau{1,2} - \mneu1 = m_{\tau}$,
while the horizontal line indicates the detector-stable life time of 100~ns.
The color coding is as in \reffi{mn1-mstau1-L}. Parameter points
in agreement with all constraints, as shown in red, are found for all
mass differences up to $\sim 30 \gev$.

For $\Delta m = \mstau{1,2} - \mneu1 > m_{\tau}$,
the staus decay promptly to $\tau \neu1$ final state with very small
lifetime $\lesssim 10^{-12}$ ns.
For smaller mass gaps, $\Delta m \lsim m_\tau$,
three-body ($\to \pi \nu_\tau \neu1$) or four-body ($\to l \nu_l \nu_\tau \neu1$)
final states dominate, which makes the stau lifetime longer, rendering
them effectively detector-stable.
It is apparent from \reffi{lifetime-delm} that the stau lifetimes in our scans
lie orders of magnitude below the age of the universe $\sim 10^{17}$~s.
This is in agreement with the non-observation of a heavy stable charged particle in
DM search experiments.
Moreover, a heavy stau of $\sim$ several hundred GeV, as obtained
in our case, is expected to have sufficiently small  number density so
as not to  hamper a successful big bang
nucleosynthesis~\cite{Jittoh:2005pq,Kawasaki:2004qu}. 
However, such long-lived staus might be subject to detector-stable
particle searches at the LHC. 
We show in \reffi{lifetime-mstau}
our data sample in the plane $\mstau{1,2}$-$\tau_{\Stau{1,2}}$.
The box shown indicates the parameter points that might be excluded by
the current searches from ATLAS for stable charged particles, based on
$36 \ifb$~\cite{Heinrich:2018pkj,ATLAS:2019gqq}, reaching up to masses
of $\mstau{} \lsim 430 \gev$. (CMS published a somewhat weaker limit
based on $\sim 13 \ifb$~\cite{CMS:2016ybj}.)
Since the ATLAS limits were obtained in a GMSB framework and the
production cross section differs slightly from our calculation
(see \refse{sec:future-pp} below) we did not exclude these points from
our data sample, but just indicate the points which may be affected
by a gray region in \reffi{lifetime-mstau}.
It should be noted that for the parameter points with the highest
masses, $\mstau{1,2} \sim 550 \gev$, there are points with a short life
time, as well as points which are detector-stable.

\begin{figure}[htb!]
\centering
\begin{subfigure}[b]{0.48\linewidth}
\centering\includegraphics[width=\textwidth]{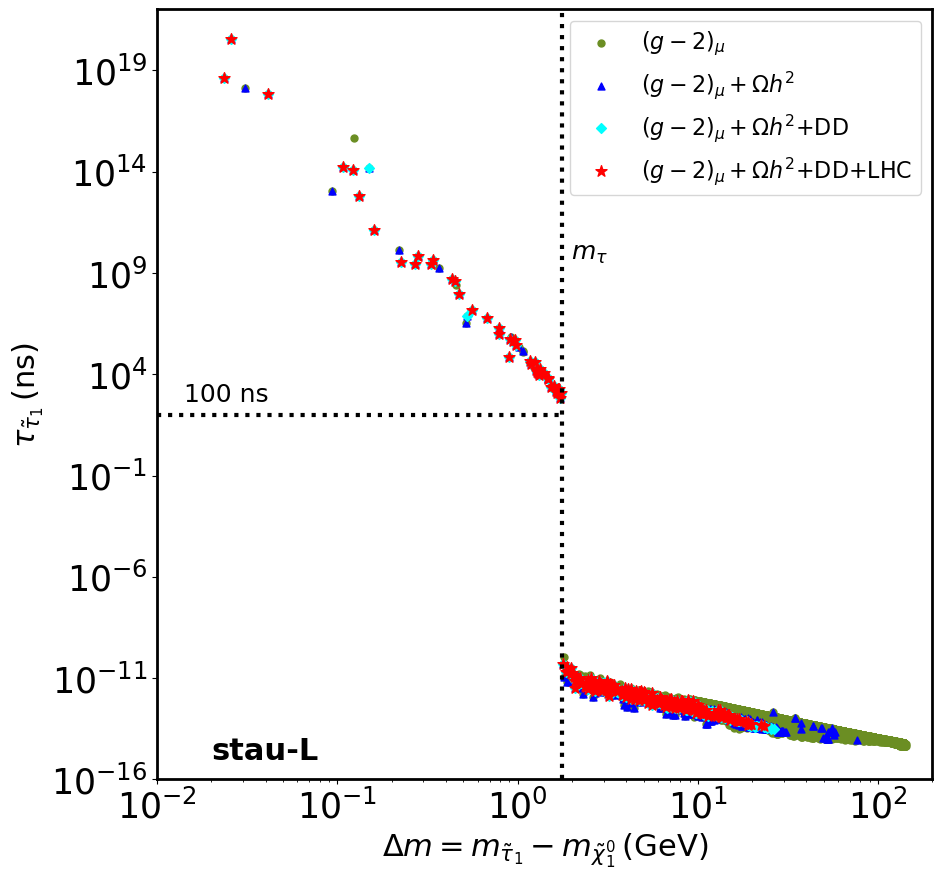}
        \caption{}
        \label{}
\end{subfigure}
~
\begin{subfigure}[b]{0.48\linewidth}
\centering\includegraphics[width=\textwidth]{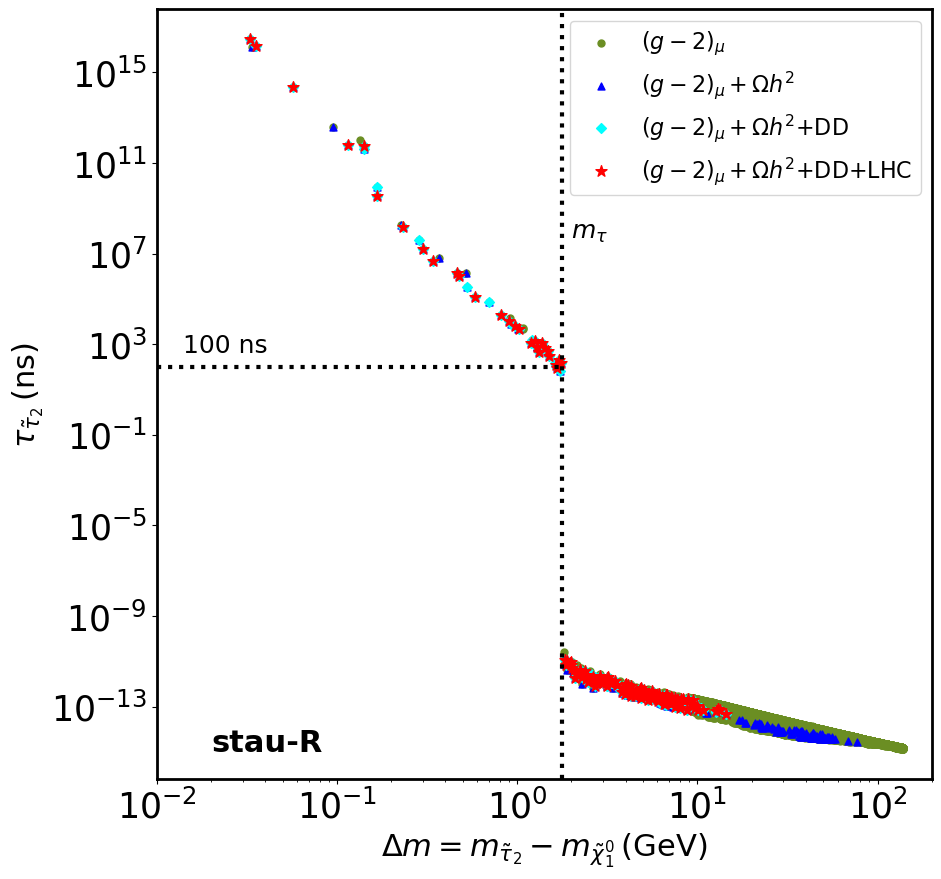}
        \caption{}
        \label{}
\end{subfigure}
\caption{Lifetime as a function of mass difference $\Delta m
        = \mstau{1,2} - \mneu1$ 
for $\Stau1$ in case-L (left) and $\Stau2$ in case-R (right).
The color coding is as in \protect\reffi{mn1-mstau1-L}.}
\label{lifetime-delm}
\end{figure}

\begin{figure}[htb!]
\centering
\begin{subfigure}[b]{0.48\linewidth}
\centering\includegraphics[width=\textwidth]{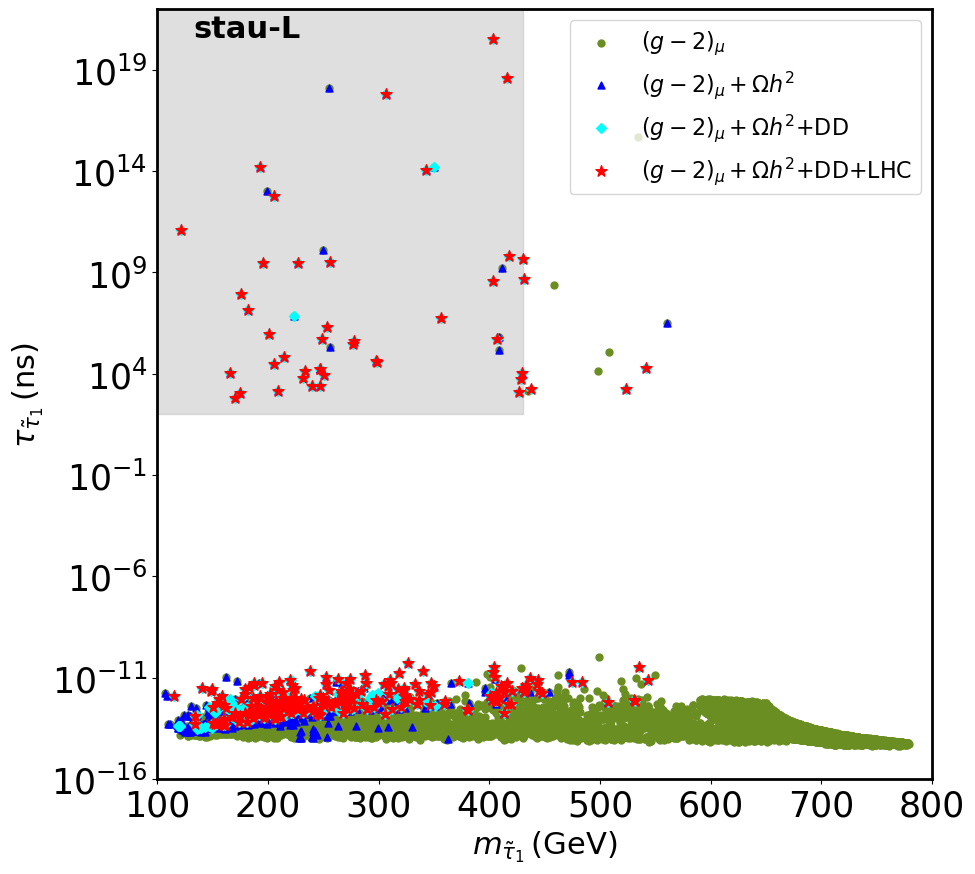}
        \caption{}
        \label{}
\end{subfigure}
~
\begin{subfigure}[b]{0.48\linewidth}
\centering\includegraphics[width=\textwidth]{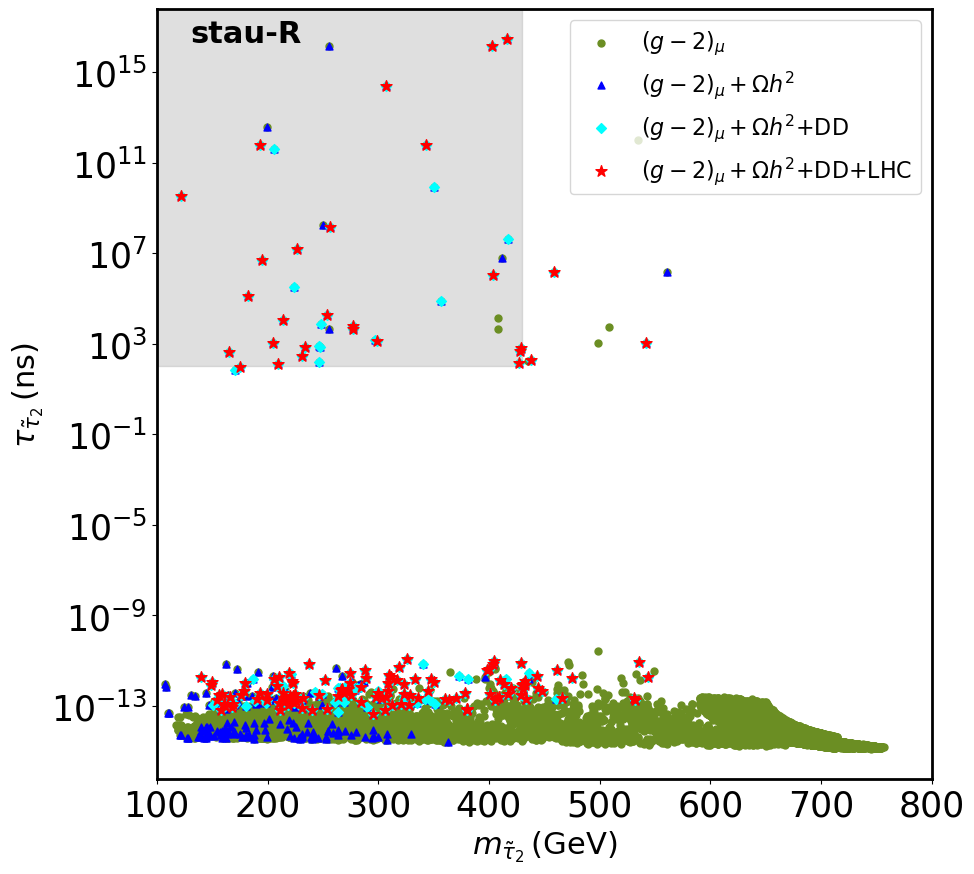}
        \caption{}
        \label{}
\end{subfigure}
\caption{Lifetime as a function of mass for $\Stau1$ in case-L (left)
and $\Stau2$ in case-R (right).
The color coding is as in \protect\reffi{mn1-mstau1-L}.}
\label{lifetime-mstau}
\end{figure}


\subsection{Prospects for DD experiments}
\label{sec:dd}

We now turn to the prospects to cover the stau-L and stau-R scenario
with DD experiments. 
In the upper (lower) plot of \reffi{ssi} we show the results of our scan
in the stau-L (stau-R) case in the $\mneu1$--$\ssi$ plane, where $\ssi$
denotes the spin-independent DM scattering cross section (which are
always substantially more relevant than the spin-dependent limits).
The color coding of the points indicates the DM relic
density, where the red points correspond to full agreement with the
Planck measurement, see \refeq{OmegaCDM}.
For the points with a lower relic density we
rescale the cross-section with a factor of ($\Omega_{\tilde \chi} h^2$/0.118)
to take into account the fact that $\neu1$ provides only a fraction of the
total DM relic density of the universe.
The various nearly horizontal lines indicate the reach of current and
future DD experiments. The current bound is given in solid blue by
LZ~\cite{LZ-new}. The future projection for LZ~\cite{LZ} and
Xenon-nT~\cite{Aprile:2020vtw} are shown as a black dashed line (which
effectively agree with each other). Furthermore, we show the projection
of the DarkSide~\cite{DarkSide} and the Argo~\cite{Argo}experiments,
which can go down to even lower cross sections, as blue dashed and blue
dot-dashed lines respectively. 
The lowest black, dot-dashed line indicates the
neutrino floor~\cite{neutrinofloor}.
Also shown for comparison is the previous ``best'' bound from
Xenon-1T~\cite{XENON}. 
The results for the stau-L and stau-R cases appear effectively
identical and we will describe them together.

\begin{figure}[htb!]
\vspace{5em}
\centering
\includegraphics[width=0.6\textwidth,trim= 0 0 180 0,clip]{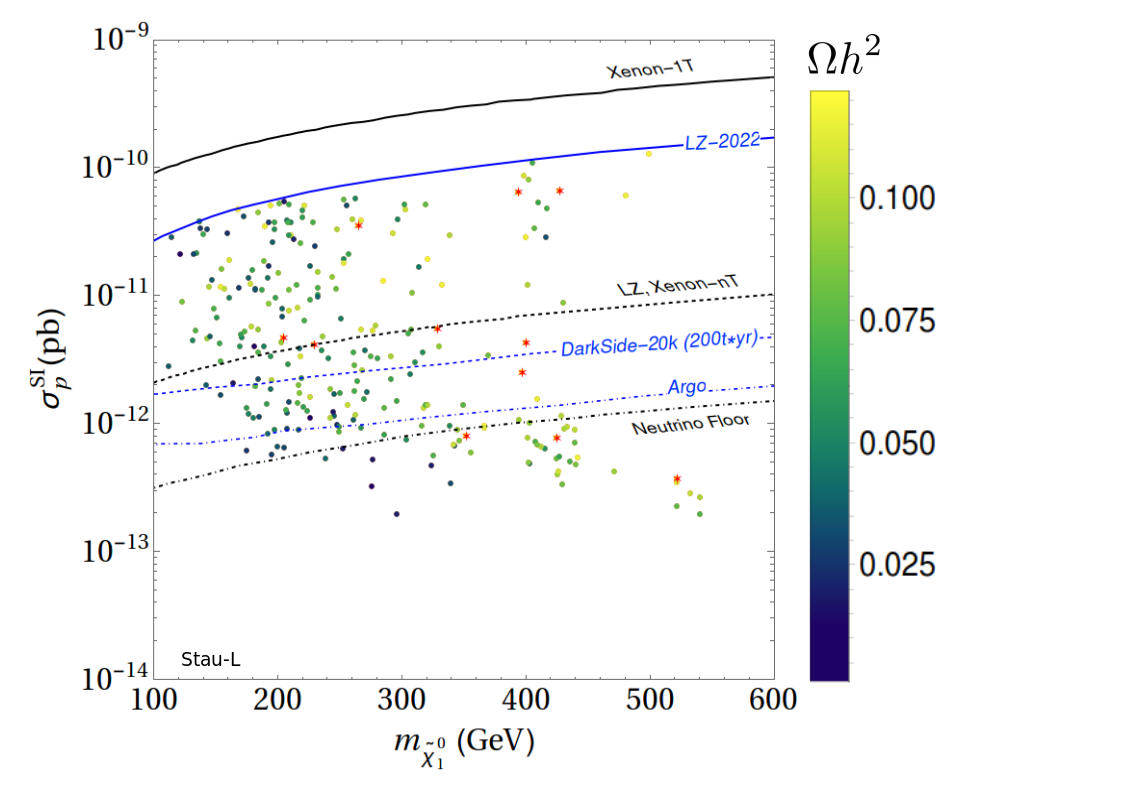}\\[2em]
\includegraphics[width=0.6\textwidth,trim=0 0 180 0,clip]{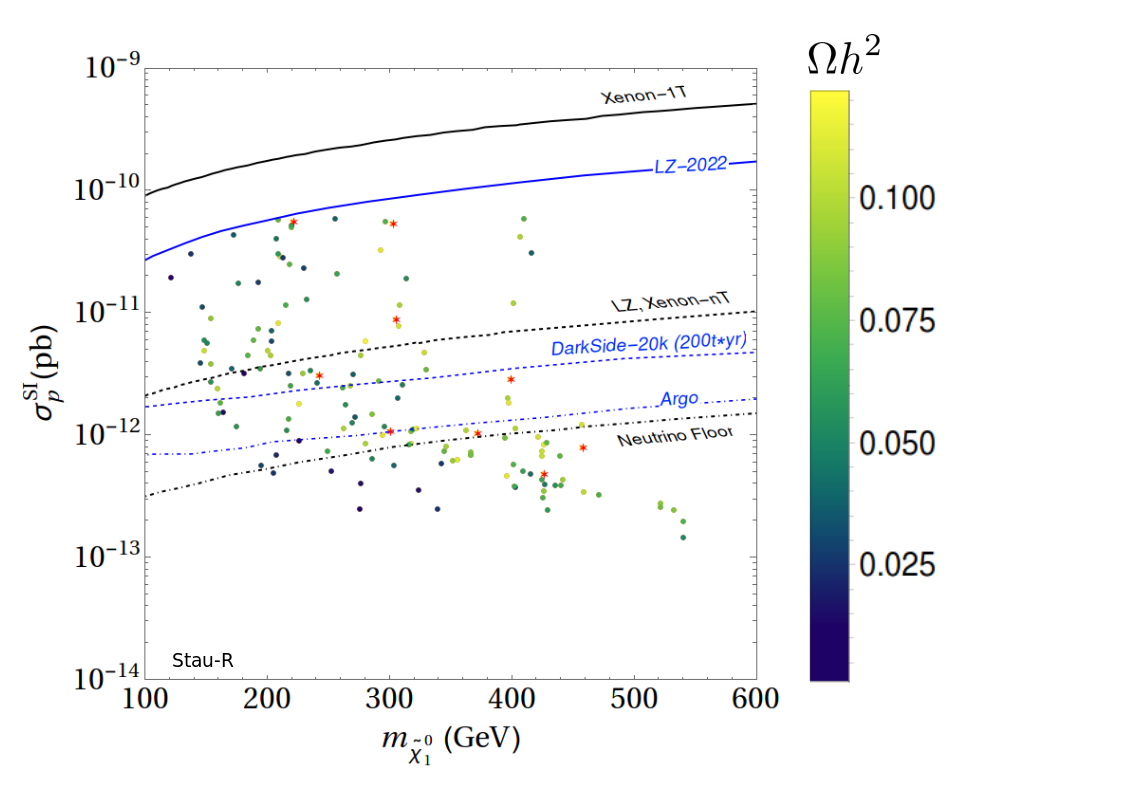}
\vspace{1em}
\caption{Surviving points $\mneu1$--$\ssi$ plane for stau-L (upper plot)
and stau-R (lower plot) case. The color coding indicates the relic
abundance, where red points denote the points with correct relic
abundance.
The various nearly horizontal lines indicate the reach of current and
future DD experiments (see text). The lowest black dot-dashed line shows the
neutrino floor.}  
\label{ssi}
\end{figure}

By construction, the upper limit of the points is provided by the
latest LZ limit (solid blue line). The points of the scans appear nearly
uniformly distributed. Also the red
points, which represent the exact relic density appear nearly uniformly
distributed. The allowed points go below the next round of Xenon based
experiments as given by LZ/Xenon-nT, but also below the Argon based
experiments, namely DarkSide and Argo. A substantial part of the allowed
parameter space is even found below the neutrino floor. Interestingly,
in both cases, stau-L and stau-R, a small population of points is found
about one order of magnitude in $\ssi$ below the neutrino floor at
masses above $\sim 500 \gev$. This is contrary to the findings for the
five scenarios listed in \refse{sec:intro}, where all points found below
the neutrino floor have $\mneu1 \lsim 500 \gev$~\cite{CHS4}.
For points below the neutrino floor the prospects to cover them in DD
experiments are currently unclear. We will discuss the prospects to
cover them at the HL-LHC and at future $e^+e^-$ colliders, i.e.\ the
complementarity of collider experiments and DD experiments, in the next
subsections. It should be remembered that the points with
masses larger than $\sim 500 \gev$ can have a short life time, or can
be detector-stable.

It should be noted here that we do not include here a discussion of
indirect detection prospections, which would go beyond the scope of our
paper. However, as we discussed in \citere{CHS4} that
the indirect detection limits only play a relevant role in the case of higgsino
DM.


\subsection{Complementarity with future collider experiments}
\label{sec:complementarity}

In this section we analyze the complementarity between future DD
experiments and collider searches. We concentrate on the parameter
points that are below the anticipated limits of LZ and XENON-nT, and in
particular on the points below the neutrino floor. 
We first show the prospects for searches for EW SUSY particles at the
approved HL-LHC~\cite{CidVidal:2018eel}. Then we explore the prospects
at possible future high-energy $e^+e^-$ colliders, such as the
ILC~\cite{ILC-TDR,LCreport} or CLIC~\cite{CLIC,LCreport}.


\subsubsection{HL-LHC prospects}
\label{sec:future-pp}

The prospects for BSM phenomenology at the HL-LHC, running at 14~TeV and
collecting 3~\iab\ of integrated luminosity per detector, have been summarized
in \citere{CidVidal:2018eel}.
For the bino DM scenario with $\Stau{}$-coannihilation, the searches that
could be the most constraining are those looking for $\Slpm$-pair
production (where one should distinguish between $\Stau{}$-pair and
$\Sel{}$+$\Smu{}$-pair production), as well as the 
$\cha1-\neu2$ production searches leading to three leptons and $\met$ in
the final state\footnote{$\chap1(\to W^+ \neu1)\cham1(\to W^-\neu1)$ production leading to two leptons and
$\met$ in the final state usually gives rise to slightly weaker bounds.}. 
So far, to our knowledge, no projected sensitivity for the former search
exists. 
Concerning the latter search, the projected 95\% C.L.\ exclusion contours were
provided by the ATLAS collaboration in \citere{CidVidal:2018eel}
for the decays $\chapm1 \neu2 \to W^\pm Z$  and $\chapm1 \neu2 \to W^\pm h$.
The corresponding limits are given for simplified model scenarios
assuming $\neu2$ and $\cha1$ to be purely wino-like and mass-degenerate
as well as $\neu1$ to be purely bino-like. 
These searches are most effective in the regions with sufficiently large
mass splitting: 
$\Delta m = \mcha1 - \mneu1 \gtrsim \MZ$ and  $\Delta m \gtrsim \Mh$
for the $W^\pm Z$  and $W^\pm h$ modes, respectively, where 
masses up to $\mcha1 = \mneu2 \sim 1.2 \tev$ can be probed. 
The parameter region where $\mcha1, \mneu2 > \mL,\mR$, the
$\chapm1, \neu2$ may also decay via sleptons of the first two
generations, weakening the above bounds (the decay via staus,
despite kinematically favored, leads to more complicated final states).
The prospect for such decay channels,
however has not been analyzed.

For stau pair production searches at the HL-LHC,
the exclusion limit can reach $690~(430) \gev$ for pure $\tilde \tau_L$
($\tilde \tau_R$) pair production, assuming a massless $\neu1$ and
a large ($\gtrsim 100 \gev$) mass difference between the stau and $\neu1$~\cite{CidVidal:2018eel}.
The discovery sensitivity for pure $\tilde \tau_L \tilde \tau_L$ production
falls in the stau mass range $110-500 \gev$. However, no discovery sensitivity
for pure $\tilde \tau_R \tilde \tau_R$ production could be obtained. This search may
be applied to constrain the pair production of the heavier stau in our scenarios.
Thus, at least a part of our parameter space may be probed by this search at the HL-LHC.

We calculated the NLO+NLL threshold resummed cross sections at the HL-LHC for
$\chap1\cham1$ and $\chap1\neu2$
production using the public package
Resummino~\cite{resummino,Bozzi:2006fw,Bozzi:2007qr,Debove:2009ia,Debove:2010kf}\footnote{We do not separately
calculate $\cham1\neu2$ production cross section as it is expected to be very similar to that of $\chap1\neu2$.}.
Despite the absence of clear future expected discovery/exclusion bounds
for the slepton searches, 
in order to provide  an estimate of the production cross section at the
HL-LHC, we also computed the corresponding cross sections for
$\tilde e_L^\pm \tilde e_L^\mp$ + 
$\tilde \mu_L^\pm \tilde \mu_L^\mp$  productions, as well as for
$\Stau{1/2}^\pm \Stau{1/2}^\mp$ production for the stau-L/R case. 

In \reffi{HLLHC-L} we present 
our results for the relevant production cross sections in the stau-L
scenario for the ``surviving'' points, i.e.\ the parameter points
passing all constraints.
In the upper row we show $\sig(pp \to \Stau{1}\Stau{1})$ as a function of
$\mstau1$ (left) and $\De m_{\Stau1} = \mstau1 - \mneu1$ (right). 
In the lower row we show $\sig(pp \to \Sel{L}\Sel{L} + \Smu{L}\Smu{L})$
(left) and $\sig(pp \to \chap1\cham1, \chap1\neu2)$ (right).
In the upper and lower left plots the green and blue stars represent
points below the sensitivity of Xenon-nT/LZ and the neutrino floor,
respectively. 
In the lower right plot the orange and green squares show the results for
$\chap1\cham1$ and $\chap1\neu2$ production for points below the
Xenon-nT/LZ sensitivity, whereas the blue and black stars indicate the
corresponding results for points below the neutrino floor.

\begin{figure}[htb!]
\vspace{2em}
\centering
\begin{subfigure}[b]{0.48\linewidth}
	\centering\includegraphics[width=\textwidth]{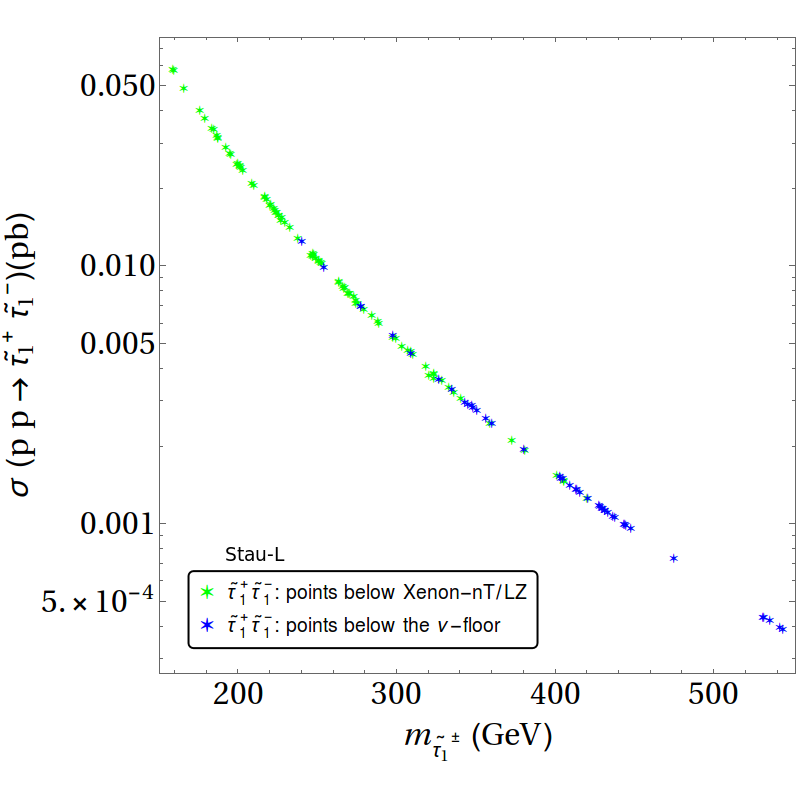}
	\caption{}
	\label{}
\end{subfigure}
~
	\begin{subfigure}[b]{0.48\linewidth}
	\centering\includegraphics[width=\textwidth]{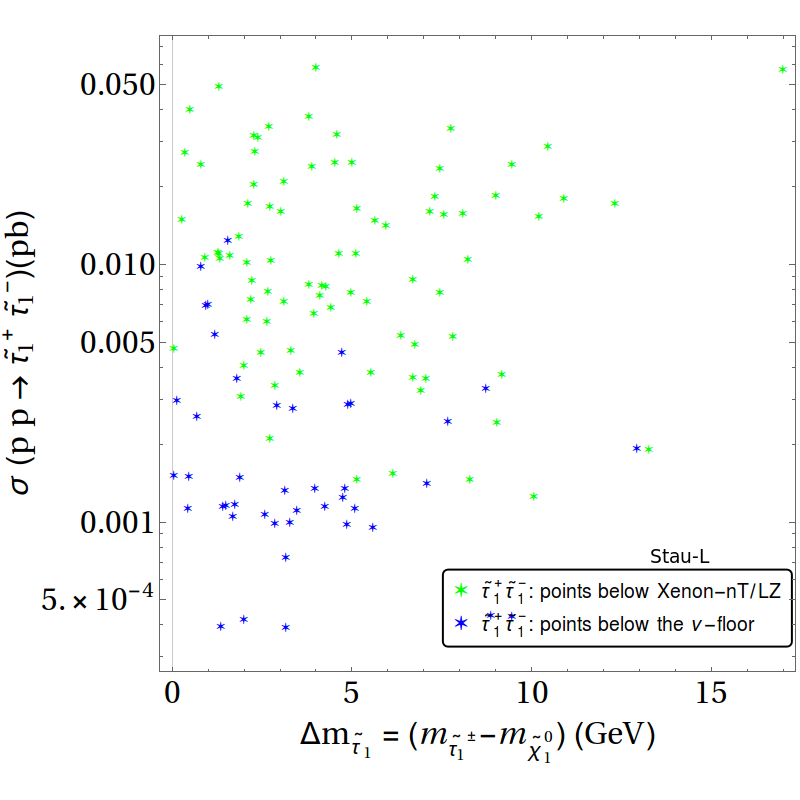}
	\caption{}
	\label{}
\end{subfigure}

\vspace{2em}
\begin{subfigure}[b]{0.48\linewidth}
	\centering\includegraphics[width=\textwidth]{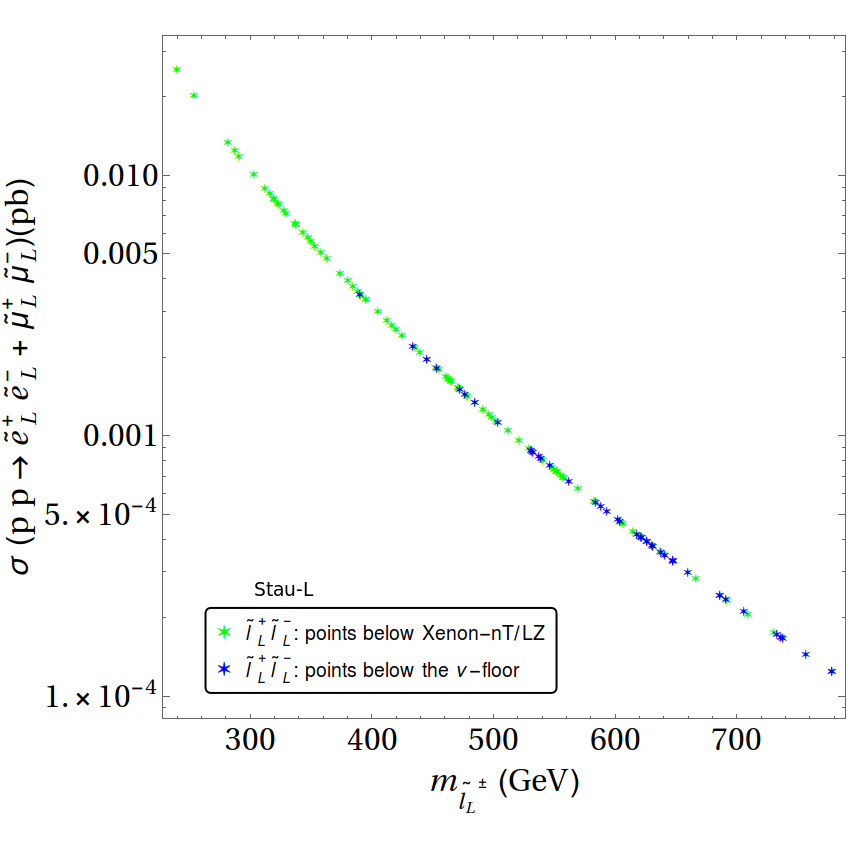}
	\caption{}
	\label{}
\end{subfigure}
~
\begin{subfigure}[b]{0.48\linewidth}
	\centering\includegraphics[width=\textwidth]{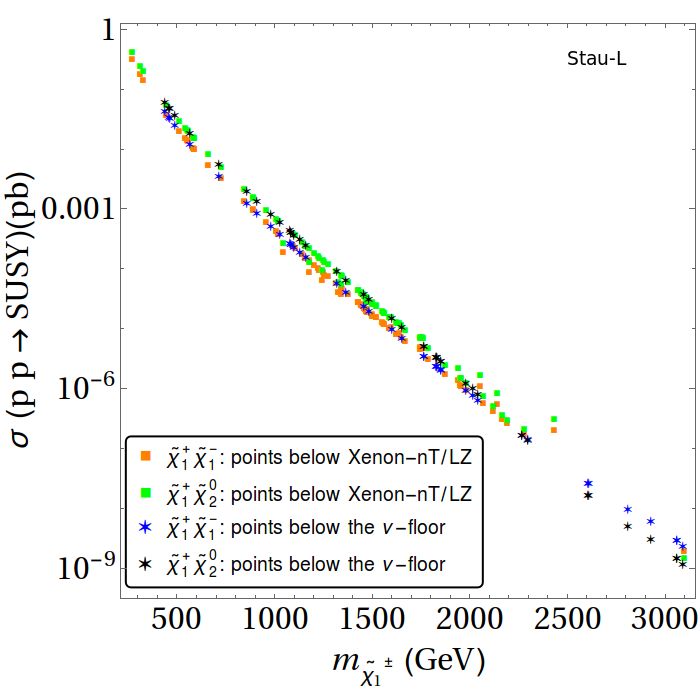}
	\caption{}
	\label{}
\end{subfigure}
\vspace{2em}
\caption{The cross-sections of the surviving parameter points in the
stau-L scenario.
Upper row: $\sig(pp \to \Stau1\Stau1)$ as a function of $\mstau1$ (left)
and $\De m_{\Stau1} = \mstau1 - \mneu1$ (right).
Lower row: $\sig(pp \to \Sel{L}\Sel{L} + \Smu{L}\Smu{L})$ (left) and
$\sig(pp \to \chap1\cham1, \chap1\neu2)$ (right).
For the color coding: see text.
}
\label{HLLHC-L}
\end{figure}

We start our discussion of the HL-LHC prospects in the stau-L scenario
with $\Stau1$ pair production, as presented in the upper row
of \reffi{HLLHC-L}. Here it should be kept in mind that these events are
characterized by compressed spectra and thus soft taus, see the
discussion in \refse{sec:staulifetime}. Concerning the numerical
results, due to phase space, the pair production cross
section goes down with $\mstau1$, starting at $50 \fb$ at
$\mstau1 \lsim 200 \gev$, going down below $0.5 \fb$ for the points with
$\mstau1 \sim 550 \gev$ (see the discussion in the previous subsection).
The latter results in less than 1500~events in each LHC experiment, and
the discovery prospects are unclear, particularly due to the compressed
spectra. For the previously discussed searches for stable charged
particles~\cite{Heinrich:2018pkj,ATLAS:2019gqq,CMS:2016ybj} no
projection for the HL-LHC is available. However, it is conceivable
that the current limits of $\mstau{} \gsim 430 \gev$ can be extended
up to $\sim 550 \gev$ with the large data sets expected from the
HL-LHC. These may exclude some of the points with $\mstau1 \sim 550 \gev$,
but not all of them, as a subset of them has a far too short life time.
As a general (but not strict) trend one observes that larger masses are
correlated with smaller $\ssi$, such that the highest mass points are
all below the neutrino floor. Consequently, for the highest mass points
both DD experiments as well as the HL-LHC do not seem to offer the
possibility to experimentally test these scenarios.

Next, in the lower left plot the cross section for slepton-pair
production of the first two generations is presented. The experimental
situation should be better than for $\Stau1$-pair production due to the
non-compressed spectra and the fact that muons and electrons are
experimentally easier accessible than taus. As discussed
in \refse{sec:stauL}, now the masses range for $\sim 300 \gev$ to
$\sim 800 \gev$. As before, the phase space leads to smaller cross
sections for larger masses, where for equal $\Smu{L}/\Sel{L}$ and
$\Stau1$ masses roughly a factor of $\sim 2$ reduction between the two production
cross sections can be observed, owing to the sum of the first and second
generation sleptons. For the largest masses now $\sim 0.1 \fb$ are
found, resulting in $\sim 300$~events, leaving the discovery prospects
very unclear.

The last cross sections discussed in the stau-L scenario are the ones
for charginos and neutralinos as shown in the lower right plot
of \reffi{HLLHC-L}. The two shown cross sections, $\chap1\champ1$ and
$\chap1\neu2$ yield effectively the same results and are thus described
together. Naturally, due do phase space the lowest masses yield the
largest cross sections, where chargino/neutralino masses around
$\sim 400 \gev$ can reach a production cross section of $\sim 1 \pb$.
However, the apparently large cross section reached for relatively light
masses has to be interpreted with caution in deriving future
exclusion/discovery potentials: on the one hand,
$\chap1, \neu2$ and $\chap1\cham1$ may decay partly via sleptons of the
first two generations, weakening the limits from gauge-boson or
Higgs-mediated decays. On the 
other hand, they may decay to some extent via $\stau$'s, relaxing
the bounds from both slepton-mediated and gauge/Higgs-boson mediated
decays.
Even more complicated are some points below the neutrino floor, which
are found with masses between $2.5 \tev$ and
$3 \tev$. From \reffi{mn1-other-L} one can see that these points
coincide with the largest $\mneu1$ values. These have production cross
sections at the level of $\sim 10^{-5} \fb$, which are clearly beyond
the reach of the HL-LHC.

\begin{figure}[htb!]
\vspace{2em}
\centering
\begin{subfigure}[b]{0.48\linewidth}
	\centering\includegraphics[width=\textwidth]{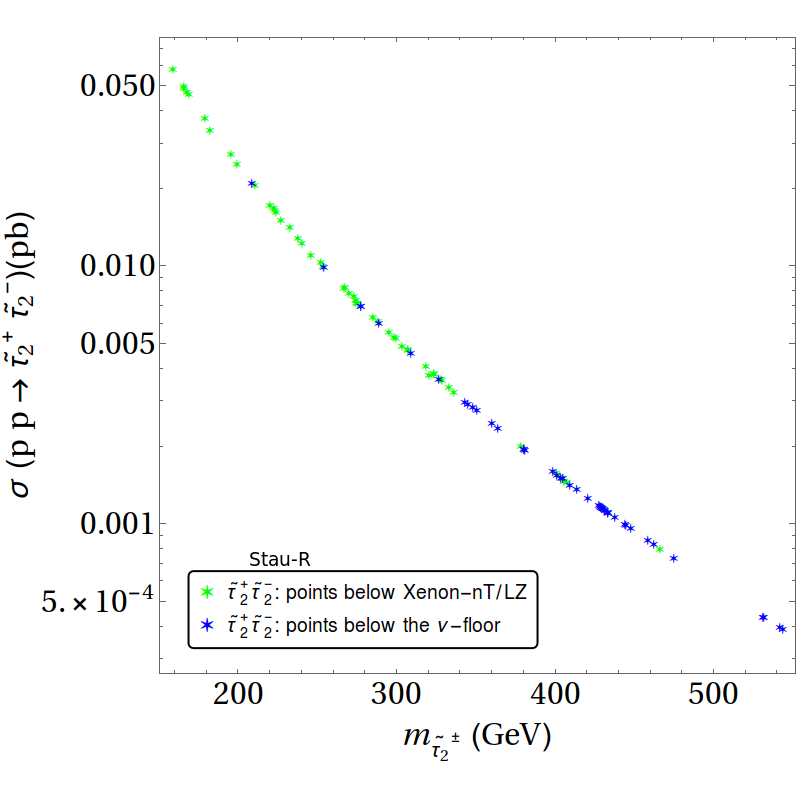}
	\caption{}
	\label{}
\end{subfigure}
~
\begin{subfigure}[b]{0.48\linewidth}
	\centering\includegraphics[width=\textwidth]{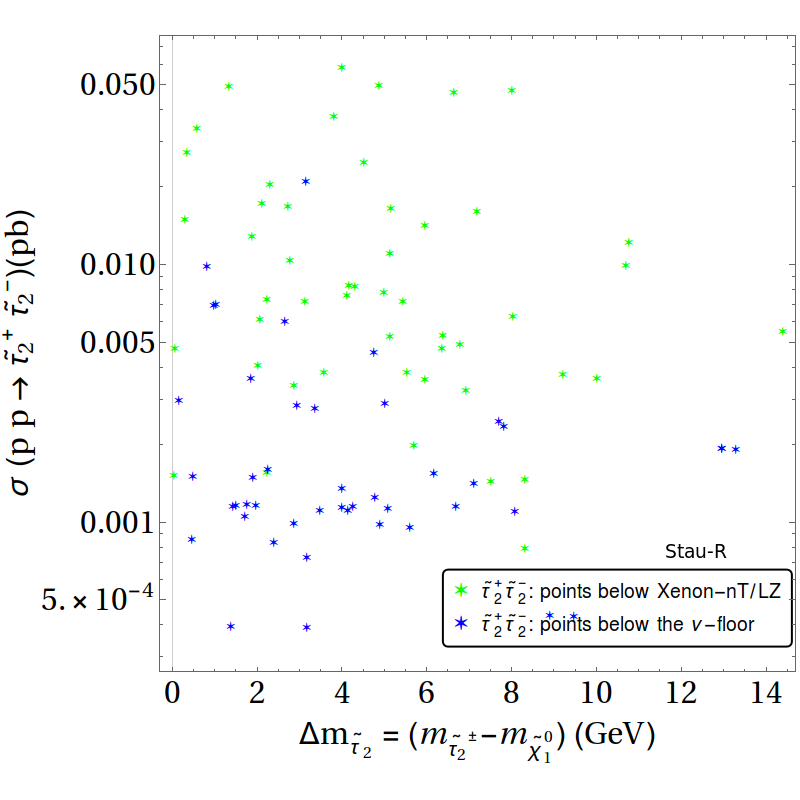}
	\caption{}
	\label{}
\end{subfigure}
~
\vspace{2em}
\begin{subfigure}[b]{0.48\linewidth}
	\centering\includegraphics[width=\textwidth]{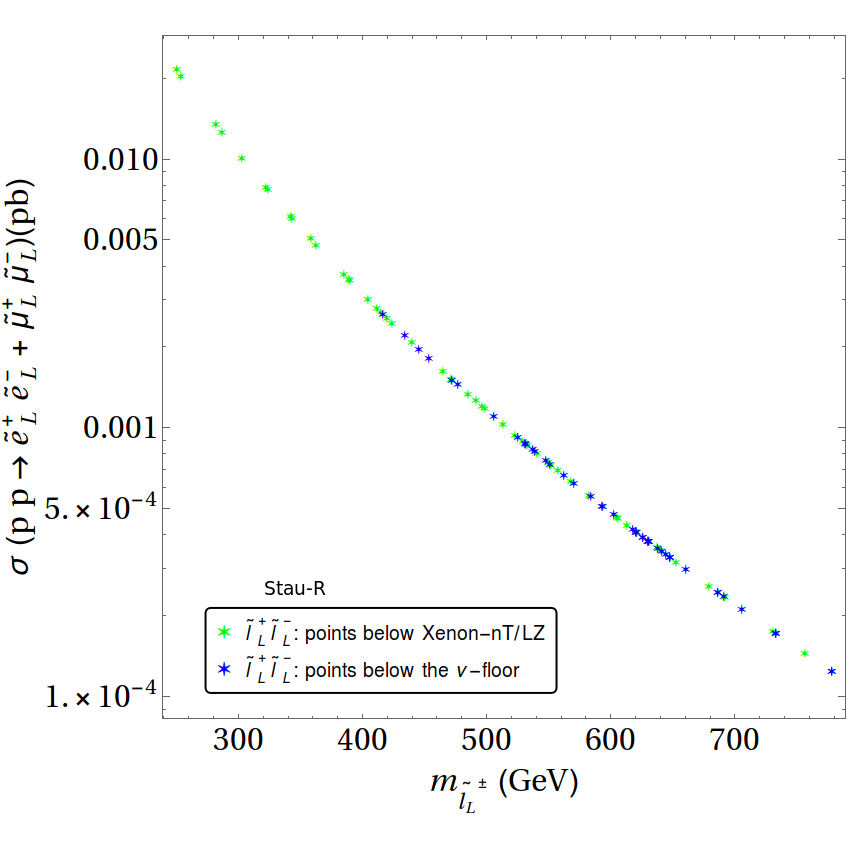}
	\caption{}
	\label{}
\end{subfigure}
~
\begin{subfigure}[b]{0.48\linewidth}
	\centering\includegraphics[width=\textwidth]{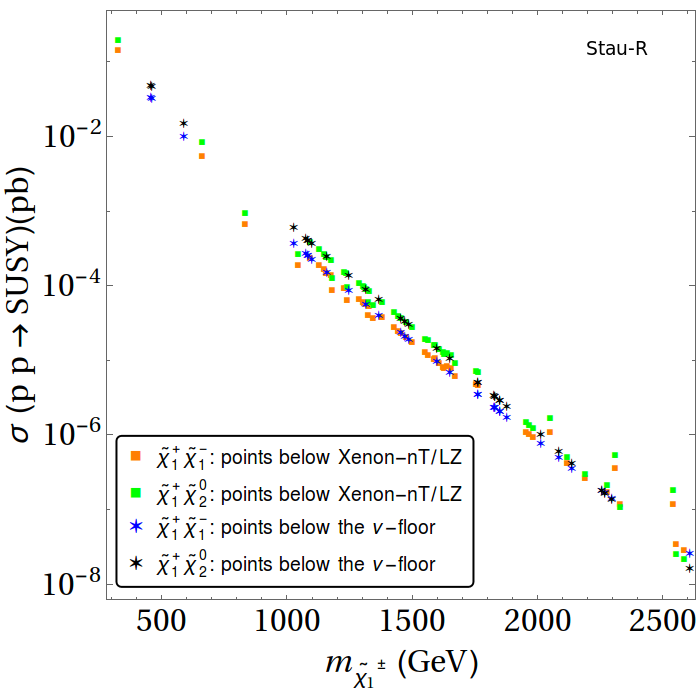}
	\caption{}
	\label{}
\end{subfigure}
\vspace{2em}
\caption{The cross-sections of the surviving parameter points in the
stau-R scenario.
Upper row: $\sig(pp \to \Stau2\Stau2)$ as a function of $\mstau2$ (left)
and $\De m_{\Stau2} = \mstau2 - \mneu1$ (right). 
Lower row: $\sig(pp \to \Sel{L}\Sel{L} + \Smu{L}\Smu{L})$ (left) and
$\sig(pp \to \chap1\cham1, \chap1\neu2)$ (right).
For the color coding: see text.}
\label{HLLHC-R}
\end{figure}

\medskip
We now turn to the $\Stau{2}$-pair, first and second generation
slepton-pair production, as well to the production of $\chap1\neu2$ and
$\chap1\cham1$ in the stau-R scenario. Our results are presented
in \reffi{HLLHC-R} with the same order of plots and the same color
coding as in \reffi{HLLHC-L}. The results turn out effectively identical
to the stau-L case. Small differences in the various production cross
sections can be observed, but they do not play any relevant
phenomenological role. Consequently, we refrain from a detailed
discussion of these EW production cross sections. However, it is
important to note that also in the stau-R case, points with
$\mneu1 \sim 550 \gev$ are found that are below the neutrino floor These points with
$\Stau{2}$-pair production cross sections of $\sim 0.5 \fb$
(where only a subset may be tested with stable charged particle
searches at the HL-LHC), first plus
second generation slepton production cross section of $\sim 0.1 \fb$ and
chargino/neutralino cross sections of $\sim 10^{-4} \fb$, 
are likely to escape the searches at the HL-LHC.

\medskip
In summary, taking into account the DD limits and the EW production
cross sections and the stable charged particle searches
discussed in this section, the complementarity
between the DD experiments and the HL-LHC can not conclusively be
answered: some points of the stau-L and stau-R scenarios could escape
both types of experiments.


\subsubsection{ILC/CLIC prospects}
\label{sec:future-ee}

The direct production of EW particles at $e^+e^-$ colliders requires a
sufficiently high center-of-mass energy, $\sqrt{s}$
that can only be reached at linear $e^+e^-$
colliders, such as ILC~\cite{ILC-TDR,LCreport} and
CLIC~\cite{CLIC,LCreport}. Those colliders can reach energies up to
$1 \tev$, and $3 \tev$, respectively.
Here we focus on the the final energy stage of the ILC, also denoted as
ILC1000. 

We evaluate the cross-sections for LSP and NLSP 
production modes for $\sqrt{s} = 1 \tev$, which can be reached at the
ILC1000.
We note here that with the anticipated higher energies at the CLIC,
larger production cross section and higher mass reach is expected.
At the ILC1000 an integrated luminosity of $8 \iab$ is
foreseen~\cite{Barklow:2015tja,Fujii:2017vwa}. 
Our cross-section predictions%
\footnote{We thank C.~Schappacher for the numerical calculations.}%
~are based on tree-level results, obtained as
in~\citeres{Heinemeyer:2017izw,Heinemeyer:2018szj}. In those articles it
was shown that the full one-loop corrections to our production cross
sections can amount up to 10-20\%\,%
\footnote{Including the full one-loop corrections here as done
in \citeres{Heinemeyer:2017izw,Heinemeyer:2018szj} would have required to
determine the preferred renormalization scheme for each
point individually
(see \citeres{Fritzsche:2013fta,Heinemeyer:2023pcc,Heinemeyer:2022apt}
for details), which goes beyond the scope of this analysis.}%
. Here we do not attempt a rigorous experimental analysis,
but rather follow the analyses
in \citeres{Berggren:2013vna,PardodeVera:2020zlr,Berggren:2020tle} that 
indicate that to a good approximation
final states with the sum of the masses smaller than the
center-of-mass energy can be detected. 

\begin{figure}[htb!]
\centering
\includegraphics[width=0.45\textwidth,height=0.45\textwidth]{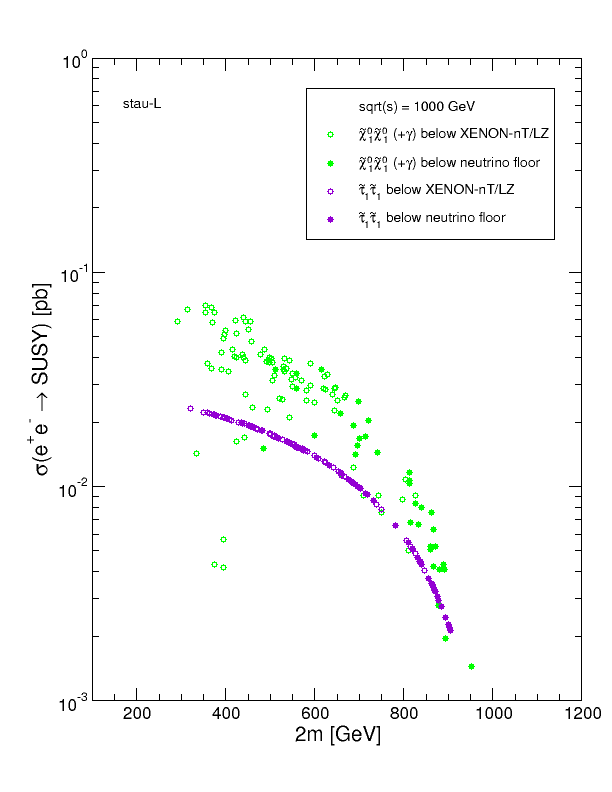}
\includegraphics[width=0.45\textwidth,height=0.45\textwidth]{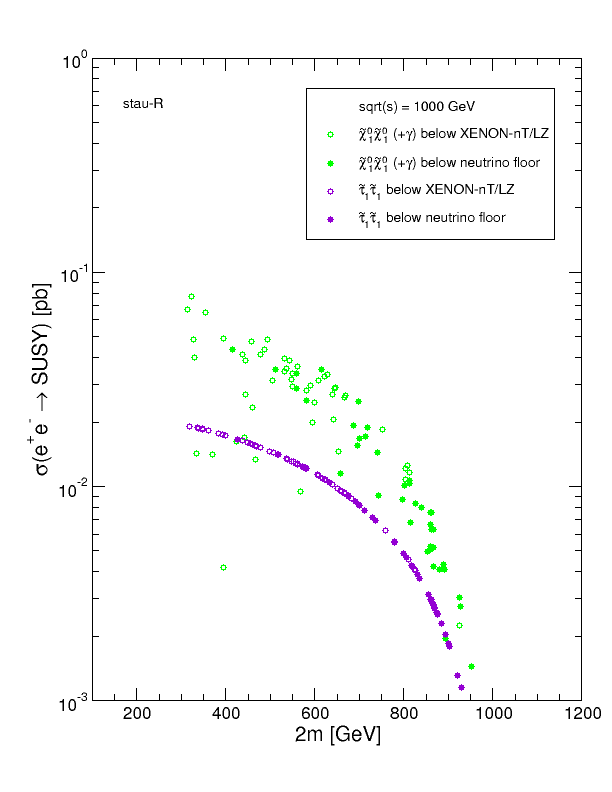}
\caption{Cross section predictions at an $e^+e^-$ collider with
$\sqrt{s} = 1000 \gev$ as a function of the sum of two final state masses.
left plot: $\Stau{1}$-coannihilation (stau-L);
right plot: $\Stau{2}$-coannihilation (stau-R).
The color code indicates the final state, open circles are below the
anticipated Xenon-nT/LZ reach, full circles are below the neutrino floor.
}
\label{fig:ee-future}
\end{figure}

In \reffi{fig:ee-future} we present the LSP and NLSP pair production cross
sections for an $e^+e^-$ collider at $\sqrt{s} = 1000 \gev$ as a
function of the two (identical) final state masses. In the left and
right plots the results for the stau-L and stau-R scenario, respectively,
are shown. In each plot we show 
$\sig(e^+e^- \to \neu1\neu1 (+ \ga))$ production%
\footnote{
Our tree level calculation does not include the photon radiation, which
appears only starting from the one-loop level. However, such an ISR
photon is crucial to detect this process due to the invisible final
state. We take our tree-level cross section as a rough approximation of
the cross section including the ISR photon, see
also \citere{Heinemeyer:2017izw} and use the notation ``$(+\ga)$''. 
}%
~in green, and $\sig(e^+e^- \to \Stau1\Stau1)$
($\sig(e^+e^- \to \Stau2\Stau2)$) in violet in the left (right) plot
of \reffi{fig:ee-future}.  

The open circles denote the points
below the anticipated XENON-nT/LZ limit, whereas the solid circles correspond to
the points below the neutrino floor. The cross sections range
roughly from $\sim 50 \fb$ for low masses to $\sim 1 \fb$ for the
largest masses shown in the plots, reaching nearly the kinematic limit
of the ILC1000, i.e.\ $\mneu1 \lsim 500 \gev$ (and the stau masses
accordingly). Assuming an integrated luminosity of $8 \iab$, this
corresponds to $\sim 8000 - 400000$ events. Consequently,
the ILC should be able to detect SUSY particles in all cases with
masses just below $500 \gev$. However, here it is important to note
that in both scenarios, stau-L and stau-R, a group of points are found
in the range of $\mneu1 \sim 550 \gev$, with the stau masses very
slightly above (where only a subset of them may be tested with
stable charged particle searches at the HL-LHC).
In order to cover these points a slightly higher
center-of-mass energy up to $\sqrt{s} \sim 1100 \gev$ would be
necessary. (A second stage CLIC with $\sqrt{s} \sim 1500 \gev$ would
clearly be sufficient to cover the two scenarios.)

These results are in contrast to the other five cases analyzed
previously in \citere{CHS4}. In these cases, and in particular for
$\Slpm$-coannihilation with all three slepton generations having
degenerate soft SUSY-breaking terms, the DD experiments together with
the ILC1000 could cover all parameter points. Allowing for a
non-degeneracy between staus and first/second generation sleptons
leads to slightly higher masses that can accommodate all constraints,
thus avoiding the detectability at the ILC1000.


\section{Conclusions}
\label{sec:conclusion}

We performed an analysis of the EW
sector of the MSSM featuring stau-coannihilation, taking into account all the latest
relevant experimental and theoretical constraints. 
The experimental results comprised the current DM relic abundance
(either as an upper limit or as a direct 
measurement), the DM direct detection (DD) experiments, the direct
searches at the LHC, and in
particular the deviation of the anomalous magnetic moment
of the muon~\cite{Abi:2021gix}. 

In our previous works, five different scenarios were
analyzed~\cite{CHS1,CHS2,CHS3,CHS4}, classified by the mechanism 
that brings the LSP relic density into agreement with the measured
values. The scenarios differ by the nature of the Next-to-LSP (NLSP), or
equivalently by the mass hierarchies between the mass scales
determining the neutralino, chargino and slepton masses.
In all scenarios a degeneracy between
the three generations of sleptons was assumed, motivated by simplicity of the analysis
as well as keeping in mind the degeneracy solution to the SUSY flavour problem. 
This becomes particularly
relevant for $\Slpm$-coannihilation where the degeneracy
results in a direct connection between the masses of the smuons involved in the explanation
for \gmin2\ to that of the  NLSP which acts as a coannihilation partner to the LSP in
explaining the DM content of the universe.
On the other hand, in
well-motivated top-down models such as mSUGRA, such degeneracy does not exist in
general and often the  
lighter stau becomes the NLSP at the EW scale via renormalization
group running, 
with the smuons/selectron masses lying slightly above.
Consequently, a full mass degeneracy of the three slepton
families should be regarded as an artificial constraint, which may lead
to artefacts in the phenomenological
analysis~\cite{deVries:2015hva,Bagnaschi:2017tru}. 

In this paper we analyzed an MSSM scenario at the EW scale,
assuming degeneracy only between smuons and selectrons, but
non-degeneracy of the staus, and enforcing stau coannihilation.
We required either the left- or the right-handed mass parameter to be
close to $M_1$ (labelled stau-L and stau-R scenarios, respectively).
For these two scenarios, corresponding more to a top-down model motivated
mass hierarchy we analyzed the viable parameter space, taking
into account all relevant theoretical and experimental constraints.
In both scenarios we find upper limits of $\mneu1 \lsim 550 \gev$ with
a very small mass difference between the neutralino LSP and the stau
NLSP. 

We have paid particular attention to the NLSP life time, as the mass
difference between the $\neu1$ and the lighter $\stau$ can be as low as
\order{10 \mev}.  Parameter points with a mass difference lower
than $m_\tau$ yield effectively detector-stable staus, which could be
subject to current searches for long lived charged particles. The
current bounds for particles with a life time larger than $\sim 100$~ns
reach up to $\mstau{} \lsim 430 \gev$ and can potentially exclude part
of these points. It is conceivable
that the current limits can be extended up to 
$\sim 550 \gev$ with the large data sets expected from the
HL-LHC. However, some points with stau masses of $\sim 550 \gev$ have
a larger mass difference and thus a far too short life time to be
affected by these searches.

In the second step we evaluated the prospects for future DD experiments
to probe the two scenarios, where we showed
explicitly the anticipated reach of XENON-nT, LZ, DarkSide and
Argo. XENON-nT and LZ have a similar reach, which is moderately improved
by DarkSide and a little further by Argo. 
For our two cases, stau-L and stau-R, the 
allowed points with the correct relic abundance can be found not only above, but
also below the projected reach of XENON-nT/LZ and the Argon based experiments.
The parameter points with the lowest DD cross section are the ones
with the larges $\neu1$ mass, nearly mass degenerate with the
respective scalar tau. These points have a DD cross section below the
neutrino floor and thus cannot be tested in with current DD
technologies. 

In continuation, we show that the HL-LHC and the ILC1000 (the $e^+e^-$
ILC operating at an energy of up to $1000 \gev$)
can play complementary roles to probe the parameter
space obtained  below the anticipated XENON-nT/LZ limit or even below
the neutrino floor.
For the HL-LHC we evaluated the production cross sections of the
NLSPs, the first/second generation of sleptons, as well as the $\cha1$
and $\neu2$. The largest cross sections are found for sleptons,
ranging from $\sim 10 \fb$ for the smallest masses to $\sim 0.1 \fb$
for the largest masses. Here it must be kept in mind that the lighter
stau is close in mass with the $\neu1$, making their observation at
the HL-LHC extremely difficult (as also discussed for the finite stau
life time above). The largest allowed stau masses, corresponding to
small mass differences will thus not be detectable at the
HL-LHC (where only part of these points may be testable with the
stable charged particle searches at the HL-LHC).
Similarly, with a production cross section of $\sim 0.1 \fb$ for
smuon/selectron production, also these EW particles will remain
elusive at the HL-LHC. The production cross sections for charginos and
neutralinos are found to be even smaller, where the largest $\neu1$
masses correspond to very large values of $\mneu2$ and
$\mcha1$. Consequently, also these searches will not be able to cover
these parameter points. 

The situation is substantially better at the ILC1000. 
It is known in general that $e^+e^-$ colliders can probe 
mass spectra with very small mass splittings.
Consequently, one does not have to rely on the production of heavier
SUSY particles, but can study the production of the LSP (with an ISR
photon) and the stau NLSP.
We have calculated the corresponding production cross sections for
all points below the XENON-nT/LZ limit or the neutrino floor.
All points within the kinematic reach have SUSY production cross
sections above $\sim 1 \fb$ and can thus be detected at the ILC1000.
However, as discussed above, 
in both scenarios, stau-L and stau-R, a group of points was found
in the range of $\mneu1 \sim 550 \gev$, with the stau masses very
slightly above. In order to cover these points a slightly higher
center-of-mass energy up to $\sqrt{s} \sim 1100 \gev$ would be
necessary. (A second stage CLIC with $\sqrt{s} \sim 1500 \gev$ would
clearly be sufficient to cover the two scenarios.)
These results are in contrast to the other five cases analyzed
previously in \citere{CHS4}, where the combination of DD experiments
and the ILC1000 covered {\it all} points below the projected
Xenon-nT/LZ limits (and thus below the neutrino floor).
The the case of $\Stau{}$-coannihilation, i.e.\ allowing for a
non-degeneracy between staus and first/second generation sleptons
leads to slightly higher masses that can accommodate all constraints,
thus avoiding the detectability at the ILC1000. 



\subsection*{Note Added}

While we were finalizing our results, the ``MUON G-2'' collaboration
published their results from Run~2 and~3~\cite{g-2-23}.
The value of \gmin2\ of these runs
\begin{align}
\amu^{\rm Run2,3} &= (11 659205.5 \pm 2.4) \times 10^{-10}~.
\label{gmt-23}
\end{align}
is well compatible with the previous results from Run~1, as well
as with the (so far) world average. The new combined value of
\begin{align}
\amu^{\rm exp-new} &= (11 659205.9 \pm 2.2) \times 10^{-10}~.
\label{gmt-exp-new}
\end{align}
compared with the SM prediction in \refeq{gmt-sm}, yields a
deviation of
\begin{align}
\Delta\amu &= (24.9 \pm 4.8) \times 10^{-10}~, 
\label{gmt-diff-new}
\end{align}
corresponding to a $5.1\sig$ discrepancy.  
While this new results will mildly shrink the
parameter ranges allowed by \gmin2, 
the overall conclusions of the paper concerning the combined upper limits on the
(N)LSP masses and future prospects remain unchanged.


\subsection*{Acknowledgments}

We thank
V.~Mitsou
and
M.A.~Sanchez Conde
for helpful discussions.
We thank C.~Schappacher for technical support.
I.S. acknowledges support from project number RF/23-24/1964/PH/NFIG/009073 and
from DST-INSPIRE, India, under (IFF) grant IFA21-PH272.
The work of S.H.\ has received financial support from the
grant PID2019-110058GB-C21 funded by
MCIN/AEI/10.13039/501100011033 and by "ERDF A way of making Europe".
MEINCOP Spain under contract PID2019-110058GB-C21 and in part by
by the grant IFT Centro de Excelencia Severo Ochoa CEX2020-001007-S
funded by MCIN/AEI/10.13039/501100011033.
We acknowledge the use of the IFT Hydra computation cluster for a part of
our numerical analysis.



\newcommand\jnl[1]{\textit{\frenchspacing #1}}
\newcommand\vol[1]{\textbf{#1}}

\newpage{\pagestyle{empty}\cleardoublepage}

\end{document}